\documentclass[letterpaper,12pt]{article}

\usepackage{array}
\usepackage{caption}
\usepackage{geometry}
\usepackage{jheppub}
\usepackage{longtable}
\usepackage{pdflscape}

\DeclareMathOperator{\Conf}{Conf}
\DeclareMathOperator{\GL}{GL}
\DeclareMathOperator{\Gr}{Gr}
\DeclareMathOperator{\SL}{SL}

\newcommand{\hel}{\textrm{k}}
\newcommand{\nkmhv}[1]{\textrm{N}^{#1}\textrm{MHV}}

\newcommand{\eqnRef}[1]{Eq.~(\ref{eqn:#1})}
\newcommand{\figRef}[1]{Fig.~\ref{fig:#1}}
\newcommand{\figRefs}[2]{Figs.~\ref{fig:#1}-\ref{fig:#2}}
\newcommand{\secRef}[1]{Sec.~\ref{sec:#1}}
\newcommand{\tabRef}[1]{Tab.~\ref{tab:#1}}

\newcommand{\ko}[1]{\mathcal{K}_{#1}}
\newcommand{\ro}[1]{\mathcal{R}_{#1}}
\newcommand{\uo}[1]{\mathcal{U}_{#1}}

\newcommand{\nProps}{n_{\textrm{props}}}
\newcommand{\nEmpty}{n_{\textrm{empty}}}
\newcommand{\nFull}{n_{\textrm{filled}}}

\title{All-Helicity Symbol Alphabets from Unwound Amplituhedra}

\author{I.~Prlina,$^1$}
\author{M.~Spradlin,$^{1,2}$}
\author{J.~Stankowicz,$^{1,3}$}
\author{S.~Stanojevic$^1$}
\author{and A.~Volovich$^{1,2}$}

\affiliation{$^1$ Department of Physics,
  Brown University,
  Providence RI 02912}

\affiliation{$^2$ School of Natural Sciences,
  Institute for Advanced Study,
  Princeton NJ 08540}

\affiliation{$^3$ Kavli Institute for Theoretical Physics,
  University of California,
  Santa Barbara CA 93106}

\emailAdd{igor\_prlina@brown.edu}
\emailAdd{marcus\_spradlin@brown.edu}
\emailAdd{james\_stankowicz@brown.edu}
\emailAdd{stefan\_stanojevic@brown.edu}
\emailAdd{anastasia\_volovich@brown.edu}

\abstract{
We review an algorithm for determining the branch points
of general amplitudes in planar $\mathcal{N}=4$ super-Yang--Mills theory from
amplituhedra.  We demonstrate how to use the recent reformulation of
amplituhedra in terms of `sign flips' in order to streamline the
application of this algorithm to amplitudes of any helicity.  In this way
we recover the known branch points of all one-loop amplitudes,
and we find an `emergent positivity' on boundaries of amplituhedra.}

\begin{document}
\maketitle

\captionsetup{width=9.09in}

\section{Introduction}

\label{sec:introduction}

Physical principles impose strong constraints on the scattering amplitudes
of elementary particles.
For example, when working at finite
order in perturbation theory,
unitarity and locality appear to constrain
amplitudes to be holomorphic functions
with poles and branch points
at precisely specified locations in the space
of complexified kinematic data describing the configuration of
particles.
Indeed, it has been a long-standing goal to understand
how to use the tightly prescribed analytic structure of
scattering amplitudes to determine them directly, without
relying on traditional (and, often computationally complex)
Feynman diagram techniques.

The connection between the physical and mathematical structure of
scattering amplitudes
has been especially well studied in
planar $\mathcal{N}=4$ super-Yang--Mills~\cite{Brink:1976bc}
SYM\footnote{We use ``SYM'' to mean the planar
limit, unless otherwise specified.}
theory in four spacetime dimensions, where the
analytic structure of amplitudes is especially tame.
The overall aim of this paper, its
predecessors~\cite{Dennen:2015bet,Dennen:2016mdk}, and its
descendant(s), is to ask a question that might be hopeless
in another, less beautiful quantum field theory:
can we understand the branch cut structure of general
scattering amplitudes in SYM theory?

The motivation for asking this question is two-fold.  The
first is the expectation that the rich mathematical
structure that underlies the integrands of SYM theory (the
rational $4L$-forms that arise from summing $L$-loop Feynman
diagrams, prior to
integrating over loop momenta)
is reflected in the corresponding scattering amplitudes.
For example, it has been observed that both
integrands~\cite{ArkaniHamed:2012nw}
and amplitudes~\cite{Golden:2013xva,Golden:2014pua,Drummond:2017ssj}
are deeply connected to the mathematics of cluster algebras.

Second, on a more practical level,
knowledge of the branch cut structure of amplitudes is the
key ingredient in the amplitude bootstrap program, which represents
the current state of the art for high loop order
amplitude calculations in SYM theory.
In particular the hexagon bootstrap (see for
example~\cite{Dixon:2014xca}), which has succeeded in
computing all six-particle amplitudes through
five loops~\cite{Caron-Huot:2016owq}, is predicated on the hypothesis
that at any loop order, these amplitudes can have branch points
only on 9 specific loci in the space of external data.
Similarly the heptagon bootstrap~\cite{Drummond:2014ffa},
which has revealed
the symbols of the seven-particle four-loop
MHV and three-loop NMHV amplitudes~\cite{Dixon:2016nkn},
assumes 42 particular branch points.
One result we hope follows from understanding the branch cut structure of
general amplitudes in SYM theory is a proof of this counting to all loop order
for six- and seven-particle amplitudes.

It is a general property
of quantum field theory (see for example~\cite{Cutkosky:1960sp,ELOP})
that the locations of singularities of
an amplitude can be determined from knowledge of the poles of
its integrand by solving the Landau
equations~\cite{Landau:1959fi}.
Constructing explicit representations for integrands can be a challenging
problem in general, but
in SYM theory this can be side-stepped by using various
on-shell methods%
~\cite{Bern:1994cg,Cachazo:2004kj,Britto:2005fq,Elvang:2013cua} to efficiently determine
the locations of integrand poles.
This problem is beautifully geometrized by
amplituhedra~\cite{Arkani-Hamed:2013jha},
which are spaces encoding representations of
integrands in such a way that the boundaries of an amplituhedron
correspond precisely to the poles of the corresponding integrand.
Therefore, as pointed out in~\cite{Dennen:2016mdk}
(which we now take as our conceptual framework),
the Landau equations can be interpreted as
defining a map that associates to any boundary of an amplituhedron
the locus in the space of external data where the corresponding
amplitude has a singularity.

Only MHV amplitudes were considered in~\cite{Dennen:2016mdk}.
In this paper we show how to extend the analysis to
amplitudes of arbitrary helicity. This is greatly aided
by a recent combinatorial reformulation of amplituhedra in terms
of ``sign flips''~\cite{Arkani-Hamed:2017vfh}.
As a specific application of our algorithm we classify
the branch points of all one-loop amplitudes
in SYM theory.
Although the singularity structure of these amplitudes
is of course well-understood (see for example%
~\cite{tHooft:1978jhc,Bern:1993kr,Bern:1994zx,Bern:1994ju,Brandhuber:2004yw,
Bern:2004ky,Britto:2004nc,Bern:2004bt,Ellis:2007qk}),
this exercise serves a useful purpose in preparing a
powerful toolbox
for the sequel~\cite{Prlina:2017tvx}
to this paper where we will see that boundaries
of one-loop amplituhedra are the basic building blocks
at all loop order.
In particular we find a surprising `emergent
positivity' on boundaries of one-loop amplituhedra that
allows boundaries to be efficiently mapped between
different helicity sectors, and recycled to higher loop levels.

The rest of this paper is organized as follows.  In~\secRef{review}
we review relevant definitions
and background material and summarize the general procedure
for finding singularities of amplitudes.
In Secs.~\ref{sec:one-loop-branches}
and~\ref{sec:one-loop-boundaries}
we classify the relevant boundaries of all one-loop amplituhedra.
Section~\ref{sec:hierarchy} outlines a simple graphical
notation for certain boundaries and shows that the one-loop
boundaries all assemble into a simple graphical hierarchy which
will prove useful for organizing higher-loop computations.
In~\secRef{momentumtwistorlandauequations} we show how to
formulate and efficiently solve the Landau equations directly in momentum
twistor space, thereby completing the identification of all branch
points of one-loop amplitudes.
The connection between these results and symbol alphabets
is discussed in~\secRef{symbology}.

\section{Review}
\label{sec:review}

This section provides a thorough introduction to the problem
our work aims to solve.
The concepts and techniques reviewed here will be
illuminated in subsequent sections via several concrete examples.

\subsection{The Kinematic Domain}
\label{sec:externaldata}

Scattering amplitudes
are (in general multivalued) functions of the kinematic
data (the energies and momenta)
describing some number of particles participating in some scattering process.
Specifically, amplitudes are functions only of the
kinematic information about the
particles entering and exiting the process,
called \emph{external data} in order to
distinguish it from information about virtual particles which may be
created and destroyed during the scattering process itself.
A general scattering amplitude in SYM theory is labeled by three integers:
the number of particles $n$, the helicity sector
$0 \le \hel \le n-4$, and the loop order $L \ge 0$, with
$L=0$ called \emph{tree level} and $L > 0$ called \emph{$L$-loop level}.
Amplitudes with $\hel = 0$
are called maximally helicity violating (MHV)
while those with $\hel > 0$ are called
(next-to-)$^{\hel}$maximally helicity violating ($\nkmhv{\hel}$).

The kinematic configuration space of SYM theory admits
a particularly simple characterization:
$n$-particle
scattering amplitudes\footnote{Here and in all that follows, we mean
components of superamplitudes suitably normalized by dividing out
the
tree-level Parke-Taylor-Nair superamplitude~\cite{Parke:1986gb,Nair:1988bq}. We expect our results to apply equally well to BDS-~\cite{Bern:2005iz}
and BDS-like~\cite{Alday:2009dv} regulated MHV
and non-MHV amplitudes.  The set of branch points of a non-MHV ratio
function~\cite{Drummond:2008vq}
should be a subset of those of the corresponding non-MHV amplitude,
but our analysis cannot exclude the possibility that it
may be a proper subset due to cancellations.}
are multivalued
functions on $\Conf_n(\mathbb{P}^3)$, the
space of configurations of $n$ points in $\mathbb{P}^3$~\cite{Golden:2013xva}.
A generic point in $\Conf_n(\mathbb{P}^3)$ may be represented by a collection
of $n$ homogeneous coordinates $Z_a^I$ on $\mathbb{P}^3$
(here $I \in \{1,\ldots,4\}$ and $a\in\{1,\ldots,n\}$) called
\emph{momentum twistors}~\cite{Hodges:2009hk}, with two such collections
considered equivalent if the corresponding $4 \times n$ matrices
$Z \equiv
(Z_1 \cdots Z_n)$ differ by left-multiplication by an element of $\GL(4)$.
We use the standard notation
\begin{align}
\label{eqn:four-bracket}
  \langle a\, b\, c\, d \rangle = \epsilon_{IJKL}
  Z_a^I Z_b^J Z_c^K Z_d^L
\end{align}
for the natural $\SL(4)$-invariant four-bracket on momentum twistors
and use the shorthand
$\langle \cdots \, \overline{a} \, \cdots \rangle =
\langle \cdots a{-}1\,a\,a{+}1
\cdots \rangle$, with the understanding that all particle labels
are always taken mod~$n$.
We write $(a\,b)$ to
denote the line in $\mathbb{P}^3$
containing $Z_a$ and $Z_b$, $(a\,b\,c)$ to denote the plane
containing $Z_a, Z_b$ and $Z_c$, and so $\overline{a}$
denotes the plane $(a{-}1\,a\,a{+}1)$.
The bar notation is motivated by \emph{parity}, which is a
$\mathbb{Z}_2$ symmetry of SYM theory that maps
$\nkmhv{\hel}$ amplitudes to
$\nkmhv{n-\hel-4}$ amplitudes while mapping the momentum twistors
according to
$\{ Z_a \} \mapsto
\{ W_a = * (a{-}1\,a\,a{+}1) \}$.

When discussing $\nkmhv{\hel}$ amplitudes it is conventional
to consider an enlarged kinematic space where the momentum
twistors are promoted to homogeneous coordinates
$\mathcal{Z}_a$, bosonized momentum twistors~\cite{Arkani-Hamed:2013jha}
on $\mathbb{P}^{\hel+3}$ which assemble into an
$n \times (\hel + 4)$ matrix $\mathcal{Z} \equiv
(\mathcal{Z}_1 \cdots \mathcal{Z}_n)$.
The analog of~\eqnRef{four-bracket} is then the $\SL(\hel+4)$-invariant
bracket which we denote by $[ \cdot ]$ instead of $\langle \cdot \rangle$.
Given some $\mathcal{Z}$ and an element of the Grassmannian $\Gr(\hel, \hel+4)$
represented by
a $\hel \times (\hel +4)$ matrix $Y$, one can obtain
an element of $\Conf_n(\mathbb{P}^3)$ by projecting onto the complement
of $Y$.
The four-brackets of the \emph{projected external data} obtained in this
way are given by
\begin{align}
\label{eqn:projection}
\langle a\, b\, c\, d \rangle \equiv [Y\, \mathcal{Z}_a\,
\mathcal{Z}_b\,\mathcal{Z}_c\,\mathcal{Z}_d]\,.
\end{align}

Tree-level amplitudes are rational functions of the brackets
while loop-level amplitudes
have both poles and branch cuts, and are properly defined
on an infinitely-sheeted cover of $\Conf_n(\mathbb{P}^3)$.
For each
$\hel$ there exists an open set
$\mathcal{D}_{n, \hel} \subset \Conf_n(\mathbb{P}^3)$
called the \emph{principal domain}
on which amplitudes are known to be holomorphic and non-singular.
Amplitudes are initially defined only on $\mathcal{D}_{n, \hel}$
and then extended to all of (the appropriate cover of) $\Conf_n(\mathbb{P}^3)$
by analytic continuation.

A simple characterization of the principal domain for $n$-particle
$\nkmhv{\hel}$ amplitudes was given
in~\cite{Arkani-Hamed:2017vfh}:
$\mathcal{D}_{n, \hel}$ may be defined as the set of points in
$\Conf_n(\mathbb{P}^3)$
that can be represented by a $Z$-matrix with the properties
\begin{enumerate}
\item $\langle a\, a{+}1 \, b\, b{+}1 \rangle > 0$ for all
$a$ and $b \not\in\{a{-}1,a,a{+}1\}$\footnote{As explained in~\cite{Arkani-Hamed:2017vfh},
the cyclic symmetry on the $n$ particle labels
is ``twisted'', which manifests itself here in the fact that if $\hel$
is even, and if $a=n$ or $b=n$, then cycling around $n$ back to 1
introduces an extra minus sign. The condition in these cases
is therefore $(-1)^{\hel+1} \langle c\, c{+}1\, n\, 1\rangle > 0$
for all $c \not\in\{1,n{-}1,n\}$.},
and
\item the sequence $\langle 1\,2\,3\,\bullet\rangle$
has precisely $\hel $ sign flips,
\end{enumerate}
where we use the notation $\bullet \in \{ 1 \,, 2 \,, \ldots \,, n \}$ so that
\begin{align}
\langle 1\, 2\, 3\, \bullet \rangle \equiv
\{ 0, 0, 0, \langle 1\, 2\, 3\, 4 \rangle,
\langle 1\, 2\, 3\, 5 \rangle, \ldots, \langle 1\, 2\, 3\, n \rangle\}\,.
\end{align}
It was also shown that an alternate but equivalent condition is to say
that
the sequence $\langle a\,a{+}1\,b\,\bullet\rangle$
has precisely $\hel$ sign flips for all $a, b$ (omitting trivial
zeros, and taking appropriate account of the twisted cyclic symmetry
where necessary).
The authors of~\cite{Arkani-Hamed:2017vfh} showed,
and we review in~\secRef{amplituhedron},
that for $Y$'s inside an N$^\hel$MHV amplituhedron,
the projected external data have the two properties above.

\subsection{Amplituhedra ...}
\label{sec:amplituhedron}

A matrix is said to be \emph{positive} or \emph{non-negative} if all of
its ordered maximal minors are positive or non-negative, respectively.
In particular, we say that the external data are positive if
the $n \times (\hel + 4)$ matrix $\mathcal{Z}$ described
in the previous section is positive.

A point in the $n$-particle $\nkmhv{\hel}$ $L$-loop \emph{amplituhedron}
$\mathcal{A}_{n,\hel,L}$ is a collection $(Y, \mathcal{L}^{(\ell)})$
consisting of a point $Y \in \Gr(\hel, \hel + 4)$
and $L$ lines $\mathcal{L}^{(1)}, \ldots, \mathcal{L}^{(L)}$
(called the \emph{loop momenta})
in the four-dimensional complement of $Y$.
We represent each $\mathcal{L}^{(\ell)}$ as a $2 \times (\hel + 4)$
matrix
with the understanding that these
are representatives of equivalence classes under the equivalence
relation that identifies any linear combination of the rows of $Y$ with zero.

For given positive external data $\mathcal{Z}$, the
amplituhedron $\mathcal{A}_{n,\hel,L}(\mathcal{Z})$
was defined in~\cite{Arkani-Hamed:2013jha} for $n \ge 4$ as the set
of $(Y, \mathcal{L}^{(\ell)})$ that can be represented as
\begin{align}
\label{eqn:ydef}
Y &= C \mathcal{Z}\,,\\
\label{eqn:dmatrixdef}
\mathcal{L}^{(\ell)} &= D^{(\ell)} \mathcal{Z}\,,
\end{align}
in terms of a $\hel \times n$ real matrix $C$ and $L$ $2 \times n$ real matrices
$D^{(\ell)}$ satisfying the positivity property that for any $0 \le m \le L$,
all $(2m+\hel) \times n$ matrices of the form
\begin{align}
\left(\begin{matrix}
D^{(i_1)} \cr
D^{(i_2)} \cr
\vdots \cr
D^{(i_m)} \cr
C
\end{matrix}\right)
\label{eqn:nonmhvpositivity}
\end{align}
are positive.  The $D$-matrices are understood as representatives of equivalence
classes and are defined only up to translations by linear combinations of rows
of the $C$-matrix.

One of the main results of~\cite{Arkani-Hamed:2017vfh} was that
amplituhedra can be characterized directly by (projected) four-brackets,
\eqnRef{projection}, without any reference to $C$ or $D^{(\ell)}$'s, by saying that for
given positive $\mathcal{Z}$, a
collection $(Y, \mathcal{L}^{(\ell)})$ lies inside
$\mathcal{A}_{n,\hel,L}(\mathcal{Z})$ if and only if
\begin{enumerate}
\item
the projected external data lie in the principal domain
$\mathcal{D}_{n,\hel}$,
\item $\langle \mathcal{L}^{(\ell)}\,a\,a{+}1\rangle > 0$
for all $\ell$ and $a$\footnote{Again, the twisted cyclic symmetry implies
that the correct condition for the case $a=n$ is
$(-1)^{\hel+1}\langle \mathcal{L}^{(\ell)}\,n\,1\rangle > 0$.},
\item for each $\ell$, the sequence $\langle
\mathcal{L}^{(\ell)}\,1\,\bullet\rangle$
has precisely $\hel + 2$ sign flips, and
\item $\langle \mathcal{L}^{(\ell_1)}\,\mathcal{L}^{(\ell_2)}\rangle > 0$
for all $\ell_1 \ne \ell_2$.
\end{enumerate}
Here the notation $\langle \mathcal{L}\, a\,b\rangle$ means
$\langle A\, B\, a\, b\rangle$ if the line $\mathcal{L}$ is represented
as $(A\,B)$ for two points $A, B$.
It was also shown that items 2 and 3 above are equivalent to saying
that the sequence $\langle \mathcal{L}^{(\ell)}\,a\,\bullet\rangle$
has precisely $\hel+2$ sign flips for any $\ell$ and $a$.

\subsection{... and their Boundaries}
\label{sec:boundaries}

The amplituhedron $\mathcal{A}_{n,\hel,L}$ is an open set with
boundaries
at loci where one or more of the inequalities in the above definitions
become saturated.  For example, there are boundaries
where $Y$ becomes
such that one or more of the projected
four-brackets $\langle a\, a{+}1\, b\, b{+}1 \rangle$ become zero.
Such projected external data lie on a boundary of the
principal domain $\mathcal{D}_{n,\hel}$.
Boundaries of this type are already present in tree-level
amplituhedra, which are well-understood and complementary to the focus
of our work.

Instead, the boundaries relevant to our analysis occur when $Y$ is
such that the projected external data are generic, but the
$\mathcal{L}^{(\ell)}$
satisfy one or more \emph{on-shell conditions} of the form
\begin{align}
\label{eqn:onshell}
\langle \mathcal{L}^{(\ell)}\,a\,a{+}1\rangle = 0 \quad
\text{and/or}  \quad
\langle \mathcal{L}^{(\ell_1)} \, \mathcal{L}^{(\ell_2)}\rangle
=0\,.
\end{align}
We refer to boundaries of this type as $\mathcal{L}$-\emph{boundaries}\footnote{In the sequel~\cite{Prlina:2017tvx} we will strengthen this definition to require that $\langle \mathcal{L}^{(1)}\,\mathcal{L}^{(2)}\rangle = 0$ at two loops.}.
The collection of loop momenta satisfying a given set of
on-shell conditions comprises a set whose connected
components we call \emph{branches}.
Consider two sets of on-shell conditions $S$, $S'$,
with  $S' \subset S$ a proper subset,
and $B$ ($B'$) a branch of solutions to $S$ ($S'$).
Since $S' \subset S$,
$B'$ imposes fewer constraints on the degrees of freedom of the loop momenta than $B$ does.
In the case when $B \subset B'$, we say $B'$ is a \emph{relaxation} of $B$.
We use $\overline{\mathcal{A}_{n,\hel,L}}$ to denote the closure
of the amplituhedron, consisting of
$\mathcal{A}_{n,\hel,L}$ together with all of its boundaries.
We say that $\mathcal{A}_{n,\hel,L}$ \emph{has a boundary
of type} $B$ if $B \cap \overline{\mathcal{A}_{n,\hel,L}} \ne \emptyset$
and $\dim(B \cap \overline{\mathcal{A}_{n,\hel,L}}) = \dim(B)$.

\subsection{The Landau Equations}
\label{sec:landauequations}

In~\cite{Dennen:2016mdk} it was argued, based on well-known and
general properties of scattering amplitudes
in quantum field theory (see in particular~\cite{Cutkosky:1960sp}),
that all information about the locations of branch points of amplitudes
in SYM theory can be extracted from knowledge
of the $\mathcal{L}$-boundaries of amplituhedra
via the Landau equations~\cite{Landau:1959fi,ELOP}.
In order to formulate the Landau equations we must parameterize
the space of loop momenta in terms of $4L$
variables $d_A$.  For example, we could take\footnote{By writing each
$\mathcal{L}$ as a $2 \times 4$ matrix, instead
of $2 \times (\hel + 4)$, we mean to imply that we
are effectively working in a gauge where
the last four columns of $Y$ are zero and so the first $\hel$ columns
of each $\mathcal{L}$ are irrelevant and do not need to be displayed.}
$\mathcal{L}^{(\ell)} = D^{(\ell)}  \mathcal{Z}$ with
\begin{align}
D^{(1)} =\left( \begin{matrix}
1 & 0 & d_1 & d_2 \cr
0 & 1 & d_3 & d_4 \end{matrix} \right), \quad
D^{(2)} =\left( \begin{matrix}
1 & 0 & d_5 & d_6 \cr
0 & 1 & d_7 & d_8 \end{matrix} \right), \quad
\text{etc.},
\end{align}
but any other parameterization works just as well.

Consider now an $\mathcal{L}$-boundary of some $\mathcal{A}_{n,\hel,L}$
on which the $L$ lines $\mathcal{L}^{(\ell)}$ satisfy
$d$ on-shell constraints
\begin{align}
f_J = 0 \qquad (J=1,2,\ldots,d)\,,
\label{eqn:landaucutequations}
\end{align}
each of which is of the form of one of the brackets shown in~\eqnRef{onshell}.
The \emph{Landau equations} for
this set of on-shell constraints
comprise~\eqnRef{landaucutequations}
together with a set of equations
on $d$ auxiliary variables $\alpha_J$ known as
\emph{Feynman parameters}:
\begin{align}
\sum_{J=1}^d \alpha_J \frac{\partial f_J}{\partial d_A} = 0\,
\qquad
(A = 1,\ldots, 4L)\,.
\label{eqn:landaukirchhoffequations}
\end{align}
The latter set of equations are sometimes
referred to as the \emph{Kirchhoff conditions}.

We are never interested in the values of the Feynman parameters,
we only want to know under what conditions
nontrivial solutions to Landau equations exist.
Here, ``nontrivial'' means that the $\alpha_J$ must not all
vanish\footnote{Solutions for
which some of the Feynman parameters vanish are often called
``subleading'' Landau singularities in the literature, in contrast to a
``leading'' Landau singularity for which all $\alpha$'s are nonzero.
We will make no use of this terminology and pay no attention
to the values of the $\alpha$'s other than ensuring they do not
all vanish.\label{footnote:subleading}}.
Altogether we have $d + 4 L$ equations in $d + 4 L$ variables
(the $d$ $\alpha_J$'s and the $4L$ $d_A$'s).  However, the Kirchhoff
conditions are clearly invariant under a projective transformation
that multiplies all of the $\alpha_J$ simultaneously by a common
nonzero number, so the effective number of free parameters
is only $d + 4 L - 1$.
Therefore, we might expect that
nontrivial solutions to the Landau equations do not generically exist, but
that they may exist on codimension-one loci in
$\Conf_n(\mathbb{P}^3)$ --- these are the loci on which the associated
scattering
amplitude may have a singularity
according to~\cite{Landau:1959fi,ELOP}.

However the structure of solutions is rather richer than this naive expectation
suggests because the equations are typically polynomial rather than linear,
and they may not always be algebraically independent.
As we will see in the examples considered
in~\secRef{momentumtwistorlandauequations},
it is common for nontrivial solutions
to exist for generic projected external data\footnote{Solutions
of this type were associated with
infrared singularities in~\cite{Dennen:2015bet}.
We do not keep track of these solutions since the infrared
structure of amplitudes in massless gauge theory is understood
to all loop order based on exponentiation~\cite{Sterman:2002qn,Bern:2005iz}.
However, if some set of Landau equations has an ``IR solution''
at some particular $\mathcal{L}^{(\ell)}$,
there may be other solutions,
at different values of $\mathcal{L}^{(\ell)}$,
that exist only
on loci of codimension one.  In such cases
we do need to keep track of the latter.}, and
it can happen that there are branches of solutions that exist only on
loci of codimension higher than one.
We will not keep track of solutions of either of these types since
they do not correspond to branch points in the space of generic projected
external
data.

There are two important points about our procedure
which were encountered in~\cite{Dennen:2016mdk} and deserve to
be emphasized.
The first is a subtlety that arises from the fact that the
on-shell
conditions satisfied on a given boundary of some
amplituhedron are not always independent.
For example, the end of Sec.~3 of~\cite{Dennen:2016mdk} discusses
a boundary of $\mathcal{A}_{n,0,2}$
described by nine on-shell conditions with the property that
the ninth is implied by the other eight.
This situation arises generically for $L > 1$,
and a procedure --- called \emph{resolution} ---
for dealing with these cases was
proposed in~\cite{Dennen:2016mdk}.
We postpone further discussion of this point to the sequel
as this paper focuses only on one-loop examples.

Second, there is a fundamental asymmetry between the two types
of Landau equations, (\ref{eqn:landaucutequations})
and (\ref{eqn:landaukirchhoffequations}), in two respects.
When solving the on-shell
conditions we are only interested
in branches of solutions that
(A1) exist for generic projected external data, and that
(A2) have nonempty intersection with $\overline{\mathcal{A}_{n,\hel,L}}$ with correct dimension.
In contrast, when further imposing the
Kirchhoff constraints on these branches,
we are interested in solutions that
(B1) exist on codimension-one loci in $\Conf_n(\mathbb{P}^3)$, and
(B2) need not remain within $\overline{\mathcal{A}_{n,\hel,L}}$.
The origin of this asymmetry was discussed in~\cite{Dennen:2016mdk}.
In brief, it
arises from Cutkoskian intuition whereby singularities of an amplitude
may arise from configurations of loop momenta
that are outside the physical domain of integration
(by virtue of being complex; or, in the current context, being outside
the closure of the amplituhedron),
and are only accessible after analytic continuation
to some higher sheet; whereas the monodromy of an amplitude around
a singularity is computed by an integral over the physical domain
with the cut propagators replaced by delta functions. The resulting
monodromy will be zero, i.e.~the branch point doesn't really exist,
if there is no overlap between the physical domain and the locus where
the cuts are satisfied, motivating (A2) above.
In summary, it is important to ``solve the on-shell conditions first''
and then impose the Kirchhoff conditions on the appropriate
branches of solutions only afterwards.

\subsection{Summary: The Algorithm}
\label{sec:summary}

The Landau equations
may be interpreted as defining a map which associates to each
boundary of the amplituhedron $\mathcal{A}_{n,\hel,L}$
a locus in $\Conf_n(\mathbb{P}^3)$ on which the corresponding
$n$-point $\nkmhv{\hel}$ $L$-loop amplitude has a singularity.
The Landau equations themselves have no way to
indicate whether a singularity is a pole or branch point.
However, it is expected that all poles in SYM theory
arise from boundaries that are present already in the
tree-level amplituhedra~\cite{Arkani-Hamed:2013jha}.
These occur when some $\langle a\,a{+}1\,b\,b{+}1\rangle$ go to
zero as discussed
at the beginning of~\secRef{boundaries}.
The aim of our work is to understand the loci where amplitudes
have branch points,
so we confine our attention to
the $\mathcal{L}$-boundaries defined in that section.

The algorithm for finding all branch points of the
$n$-particle $\nkmhv{\hel}$ $L$-loop amplitude is therefore simple
in principle:
\begin{enumerate}
\item Enumerate all $\mathcal{L}$-boundaries of $\mathcal{A}_{n,\hel,L}$ for
generic projected external data.
\item For each $\mathcal{L}$-boundary, identify the codimension-one loci (if there are any)
in $\Conf_n(\mathbb{P}^3)$ on which the corresponding Landau
equations admit nontrivial solutions.
\end{enumerate}

However, it remains a difficult and important outstanding problem
to fully characterize the boundaries of general amplituhedra.
In the remainder of this paper we focus on the special case $L=1$,
since all $\mathcal{L}$-boundaries of $\mathcal{A}_{n,\hel,1}$
(which have been discussed extensively in~\cite{Bai:2015qoa})
may be enumerated directly for
any given $n$:
\begin{enumerate}
\item[1(a).]{Start with a list of all possible sets of on-shell conditions
of the form
$\langle\mathcal{L}\,a\,a{+}1\rangle = 0$.}
\item[1(b).]{For each such set, identify all branches of solutions that exist
for generic projected external data.}
\item[1(c).]{For each such branch $B$, determine the values of $\hel$ for
which $\mathcal{A}_{n,\hel,1}$ has a boundary of type $B$.}
\end{enumerate}
It would be enormously inefficient to carry out
this simple-minded
algorithm beyond one loop.  Fortunately, we will see in the sequel
that the one-loop results of this paper can be exploited very
effectively to generate $\mathcal{L}$-boundaries of $L>1$
amplituhedra.

\section{One-Loop Branches}
\label{sec:one-loop-branches}

In this section we carry out steps 1(a) and 1(b) listed at the
end of~\secRef{summary}.
To that end we first introduce a graphical
notation for representing sets of on-shell conditions
via \emph{Landau diagrams}.
Landau diagrams take the form of ordinary
Feynman
diagrams,
with external lines labeled $1, \ldots, n$ in cyclic order
and one internal line (called a \emph{propagator})
corresponding to each on-shell condition.
Landau diagrams relevant to amplituhedra are always planar.
Each internal
face of an $L$-loop Landau diagram
is labeled by a distinct $\ell \in \{1, \ldots, L\}$,
and each external face may be labeled by the pair $(a\,a{+}1)$ of
external lines bounding that face.

The set of on-shell conditions encoded in
a given Landau diagram is read off as follows:
\begin{itemize}
\item
To each propagator bounding an internal face
$\ell$ and an external face $(a\,a{+}1)$ we associate
the on-shell condition
$\langle \mathcal{L}^{(\ell)}\,a\,a{+}1\rangle = 0$.
\item To each propagator bounding two internal
faces $\ell_1$, $\ell_2$
we associate the on-shell
condition $\langle \mathcal{L}^{(\ell_1)} \, \mathcal{L}^{(\ell_2)}\rangle
= 0$.
\end{itemize}

At one loop we only have on-shell conditions of the first type.
Moreover, since $\mathcal{L}$ only has four degrees of freedom
(the dimension of $\Gr(2,4)$ is four), solutions to a set of
on-shell conditions will exist for generic projected external
data only if the number of conditions is $d \le 4$.
Diagrams with $d=1,2,3,4$ are respectively named tadpoles, bubbles,
triangles and boxes.
The structure of solutions to a set of on-shell conditions can
change significantly depending on how many pairs of conditions
involve adjacent indices.
Out of abundance of caution it is therefore necessary to consider
separately the eleven distinct types of Landau diagrams shown
in the second column of~\tabRef{bigtable}.
For $d>1$ their names are qualified
by indicating the number of nodes with valence
greater than three, called \emph{masses}.
These rules suffice to uniquely name each distinct type of diagram
except the two two-mass boxes shown in~\tabRef{bigtable}
which are conventionally called
``easy'' and ``hard''. This satisfies step 1(a) of the algorithm.

Proceeding now to step 1(b), we display in the third column
of~\tabRef{bigtable} all branches of
solutions (as always, for generic projected external data) to the
on-shell conditions associated to each Landau diagram.
These expressions are easily checked by inspection or by a short
calculation.  More details and further discussion of the
geometry of these problems can be found
for example in~\cite{ArkaniHamed:2010gh}.
The three-mass triangle solution involves the quantities
\begin{equation}
\begin{aligned}
\rho(\alpha)&=- \alpha \langle  i\, j{+}1  \, k\, k{+}1\rangle-
(1-\alpha)\langle  i{+}1 \, j{+}1 \, k\, k{+}1\rangle \,,
\cr
\sigma(\alpha)&=\alpha\langle  i \, j \, k\, k{+}1\rangle +
(1-\alpha)\langle   i{+}1\,j \, k\, k{+}1\rangle \,,
\label{eqn:rhosigma}
\end{aligned}
\end{equation}
and the four-mass box solution is sufficiently messy that we have
chosen not to write it out explicitly.

Altogether there are nineteen distinct types of branches, which we
have numbered (1) through (19) in~\tabRef{bigtable}
for ease of reference.
The set of solutions to any set of on-shell
conditions of the form $\langle \mathcal{L} \, a \, a{+}1 \rangle$
must be closed under parity, since each line
$(a \, a{+}1)$ maps to itself.
Most sets of on-shell conditions have two branches of
solutions related to each other by parity.
Only the tadpole, two-mass bubble, and three-mass triangle
(branches (1), (4), and (9) respectively)
have single branches of solutions that are closed under parity.

\newgeometry{left=1.4in,right=-0.6in,bottom=0.5in}
\begin{landscape}
\setlength{\topmargin}{0.25in}
\setlength{\footskip}{0.25in}
\begin{longtable}[1]
{
>{\centering\arraybackslash} m{0.10\textwidth}
>{\centering\arraybackslash} m{0.16\textwidth}
>{\arraybackslash} m{0.32\textwidth}
>{\centering\arraybackslash} m{0.16\textwidth}
>{\centering\arraybackslash} m{0.09\textwidth}
>{\centering\arraybackslash} m{0.25\textwidth}
}
Name & Landau Diagram & Branches
& $\hel$-Validity & Low-$\hel$
Twistor Diagram & $\begin{array}{c}{\mbox{Singularity}}\\
{\mbox{Locus/Loci}}
\end{array}$
\\
\hline\hline
tadpole
($n \ge 4$)
&
\includegraphics{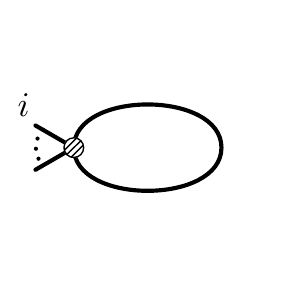}
&
$\begin{array}{rcl}
(1)~\mathcal{L} &=& (\alpha Z_i + (1{-}\alpha) Z_{i+1}, A)
\end{array}$
&
$\begin{array}{rcl}
\hphantom{n{-}}0 \le \hel \le n{-}4
\end{array}$
&
\includegraphics{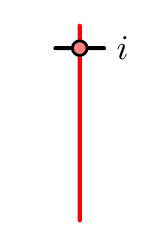}
&
$0$
\\[-15pt]
one-mass bubble
($n \ge 4$)
&
\includegraphics{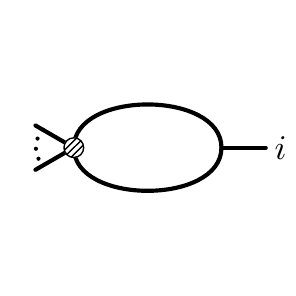}
&
$\begin{array}{rcl}
(2)~\mathcal{L} &=& (Z_i, A) \\ [10pt]
(3)~\mathcal{L} &=& \overline{i} \cap P
\end{array}$
&
$\begin{array}{rcl}
0 & \le \hel \le & n{-}4\\ [10pt]
n{-}4 & \ge \hel \ge & 0
\end{array}$
&
\includegraphics{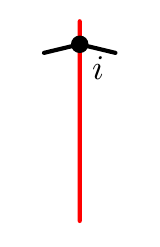}
&
$0$
\\[-15pt]
two-mass bubble
($n \ge 4$)
&
\includegraphics{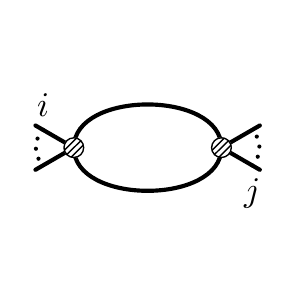}
&
$\begin{array}{rcl}
(4)~\mathcal{L} &=&(\alpha Z_i + (1{-}\alpha) Z_{i+1},
\\
&&\hphantom{(}
\beta Z_j + (1{-}\beta) Z_{j+1})
\end{array}$
&
$\begin{array}{rcl}
\hphantom{n{-}}0 \le \hel \le n{-}4
\end{array}$
&
\includegraphics{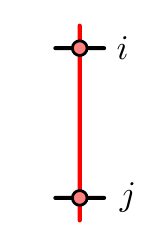}
&
$\langle i\,i{+}1\,j\,j{+}1\rangle$
\\[-15pt]
one-mass triangle
($n \ge 4$)
&
\includegraphics{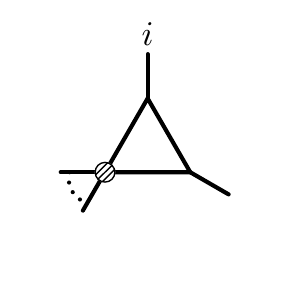}
&
$\begin{array}{rcl}
(5)~\mathcal{L} &=& (Z_i, \alpha Z_{i+1} + (1{-}\alpha) Z_{i+2}) \\ [10pt]
(6)~\mathcal{L} &=& (Z_{i+1}, \alpha Z_{i-1} + (1{-}\alpha) Z_i)
\end{array}$
&
$\begin{array}{rcl}
0 & \le \hel \le & n{-}4\\ [10pt]
n{-}4 & \ge \hel \ge & 0
\end{array}$
&
\includegraphics{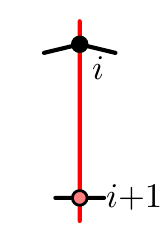}
&
$0$
\\[-15pt]
two-mass triangle
($n \ge 5$)
&
\includegraphics{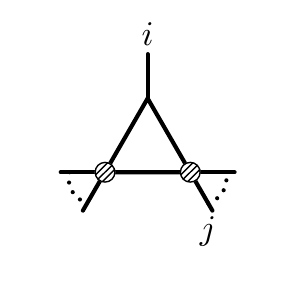}
&
$\begin{array}{rcl}
(7)~\mathcal{L} &=& (Z_i, \alpha Z_j + (1{-}\alpha) Z_{j+1})\\ [10pt]
(8)~\mathcal{L} &=& \overline{i} \cap (j\, j{+}1\, A)
\end{array}$
&
$\begin{array}{rcl}
0 & \le \hel \le & n{-}5\\ [10pt]
n{-}4 & \ge \hel \ge & 1
\end{array}$
&
\includegraphics{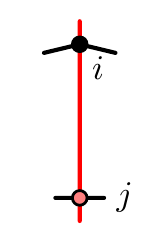}
&
$0$
\\[-15pt]
three-mass triangle
($n \ge 6$)
&
\includegraphics{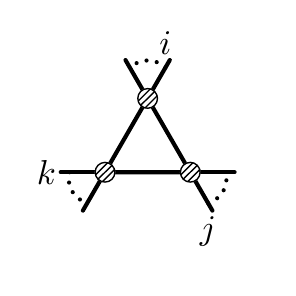}
&
$\begin{array}{rcl}
(9)~\mathcal{L} &=& (\alpha Z_i + (1{-}\alpha) Z_{i+1},\\
&&\hphantom{(} \rho(\alpha)
Z_j + \sigma(\alpha) Z_{j+1})
\end{array}$
&
$\begin{array}{rcl}
\hphantom{n{-}}1 \le \hel \le n{-}5
\end{array}$
&
\includegraphics{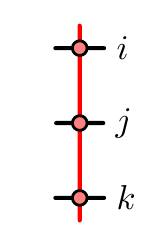}
&
$f_{ij} f_{jk} f_{ki}$
\\
\pagebreak
Name & Landau Diagram & Branches
& $\hel$-Validity &
Low-$\hel$ Twistor Diagram & $\begin{array}{c}{\mbox{Singularity}}\\
{\mbox{Locus/Loci}}
\end{array}$
\\
\hline\hline
one-mass box
($n \ge 5$)
&
\includegraphics{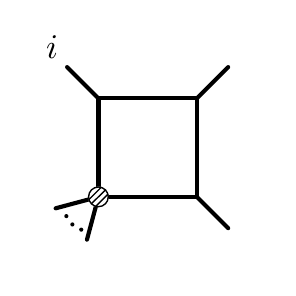}
&
$\begin{array}{rcl}
(10)~\mathcal{L} &=& (i\, i{+}2) \\ [10pt]
(11)~\mathcal{L} &=& \overline{i} \cap \overline{i{+}2}
\end{array}$
&
$\begin{array}{rcl}
0 & \le \hel \le & n{-}5\\ [10pt]
n{-}4 & \ge \hel \ge & 1
\end{array}$
&
\includegraphics{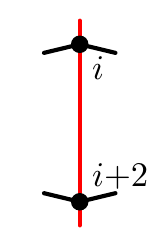}
&
$\langle i\, \overline{i{+}2}\rangle \langle \overline{i}\, i{+}2\rangle$
\\[-15pt]
two-mass easy box
($n \ge 6$)
&
\includegraphics{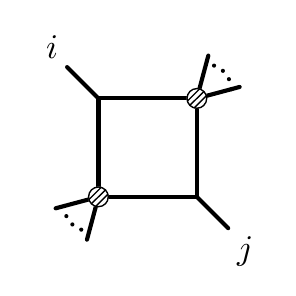}
&
$\begin{array}{rcl}
(12)~\mathcal{L} &=& (i\, j) \\ [10pt]
(13)~\mathcal{L} &=& \overline{i} \cap \overline{j}
\end{array}$
&
$\begin{array}{rcl}
0 & \le \hel \le & n{-}6\\ [10pt]
n{-}4 & \ge \hel \ge & 2
\end{array}$
&
\includegraphics{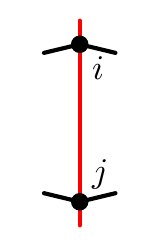}
&
$\langle i\, \overline{j} \rangle \langle \overline{i}\,j\rangle$
\\[-15pt]
two-mass hard box
($n \ge 6$)
&
\includegraphics{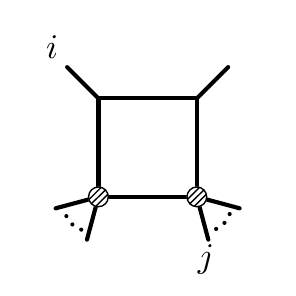}
&
$\begin{array}{rcl}
(14)~\mathcal{L} &=& \overline{i{+}1} \cap (i\,j\,j{+}1) \\ [10pt]
(15)~\mathcal{L} &=& \overline{i} \cap (i{+}1\,j\,j{+}1)
\end{array}$
&
$\begin{array}{rcl}
1 & \le \hel \le & n{-}5\\ [10pt]
n{-}5 & \ge \hel \ge & 1
\end{array}$
&
\includegraphics{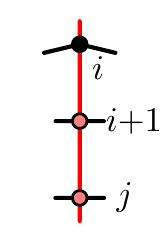}
&
$\langle \overline{i}\,i{+}2\rangle \langle i\,i{+}1\,j\,j{+}1\rangle$
\\[-15pt]
three-mass box
($n \ge 7$)
&
\includegraphics{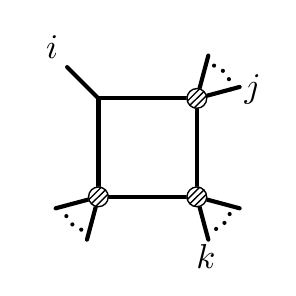}
&
$\begin{array}{l}
(16)~\mathcal{L} = (i\,j\,j{+}1) \cap (i\,k\,k{+}1) \\ [10pt]
(17)~\mathcal{L} = ( \overline{i} \cap (j\, j{+}1), \overline{i}
\cap (k\,k{+}1))
\end{array}$
&
$\begin{array}{rcl}
1 & \le \hel \le & n{-}6\\ [10pt]
n{-}5 & \ge \hel \ge & 2
\end{array}$
&
\includegraphics{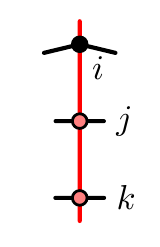}
&
$\langle i(i{-}1\,i{+}1)(j\,j{+}1)(k\,k{+}1)\rangle$
\\[-15pt]
four-mass box
($n \ge 8$)
&
\includegraphics{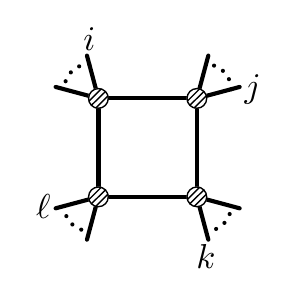}
&
$
\left.\begin{aligned}
(18)~\mathcal{L} &=&\\
(19)~\mathcal{L} &=&
\end{aligned}\right\}
~ \mbox{see Tab.~2 of~\cite{Bourjaily:2013mma}}$
&
$\begin{array}{rcl}
2 & \le \hel \le & n{-}6\\ [10pt]
n{-}6 & \ge \hel \ge & 2
\end{array}$
&
\includegraphics{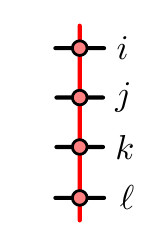}
&
$\begin{array}{r}
(f_{ij} f_{k\ell} - f_{ik} f_{j\ell} + f_{i\ell} f_{jk})^2
\\
-4 f_{ij} f_{jk} f_{k\ell} f_{i\ell}
\equiv \Delta_{ijk\ell}
\end{array}$
\\[-15pt]
\caption[One-loop Landau diagrams, branches, twistor diagrams, and loci.]{%
This table shows: the eleven Landau diagrams corresponding
to sets of one-loop on-shell conditions that can be satisfied for
generic projected
external data; the nineteen branches of solutions
to these on-shell conditions; the range of $\hel$ for which
$\nkmhv{\hel}$ amplituhedra have boundaries of each type;
the twistor diagram depicting the low-$\hel$ solution (or one
low-$\hel$ solution for the one-mass triangle and two-mass hard box);
the loci in
$\Conf_n(\mathbb{P}^3)$ where the Landau equations for each branch
admit nontrivial solutions (where the
quantity in the last column vanishes).
At one loop it happens that the loci are the same for each branch
of solutions to a given set of on-shell conditions.
Here $\alpha, \beta$ are arbitrary numbers, $A$ is an arbitrary
point in $\mathbb{P}^3$, $P$ is an arbitrary plane in $\mathbb{P}^3$,
$\rho(\alpha), \sigma(\alpha)$ are defined in~\eqnRef{rhosigma},
$f_{ab} \equiv \langle a\,a{+}1\,b\,b{+}1\rangle$,
and $\langle i(i{-}1\,i{+}1)(j\,j{+}1)(k\,k{+}1)\rangle
\equiv \langle i{-}1\,i\,j\,j\,{+}1\rangle \langle i\,i{+}1\,k\,k{+}1\rangle
- (j \leftrightarrow k)$.
}
\label{tab:bigtable}
\end{longtable}
\end{landscape}
\restoregeometry

\section{One-Loop Boundaries}
\label{sec:one-loop-boundaries}

We now turn to the last step 1(c) from the end
of~\secRef{summary}: for each of the nineteen branches
$B$ listed in~\tabRef{bigtable},
we must determine the values of $\hel$
for which $\mathcal{A}_{n,\hel,1}$ has a boundary of type $B$
(defined in~\secRef{boundaries}).
The results of this analysis are listed in the fourth column
of the \tabRef{bigtable}.  Our strategy for obtaining these results is two-fold.

In order to prove that an amplituhedron has a boundary of type $B$,
it suffices to write down a pair of matrices $C, D$ such
that definitions~(\ref{eqn:ydef}) and~(\ref{eqn:dmatrixdef})
hold,
$C$ and
$\left( \begin{smallmatrix} D \\ C \end{smallmatrix} \right)$
are both non-negative,
and the external data projected through
$Y = C \mathcal{Z}$ are generic for generic positive $\mathcal{Z}$.
We call such a pair $C, D$ a \emph{valid configuration} for $B$.
In the sections below we present explicit valid configurations
for each
of the nineteen branches.
Initially we consider for each branch only the lowest value of $\hel$
for which a valid configuration exists;
in~\secRef{largek} we explain how to grow
these to larger values of $\hel$
and establish the upper bounds on $\hel$ shown in \tabRef{bigtable}.

However, in order to prove that an amplituhedron does not have a boundary
of type $B$, it does not suffice to find a configuration that is not valid;
one must show that no valid configuration exists.
We address this problem in the next section.

\subsection{A Criterion for Establishing Absent Branches}

Fortunately, for $\mathcal{L}$-boundaries of the type under consideration
there is a simple criterion for establishing when no valid configuration
can exist.
The crucial ingredient is that if $(Y,\mathcal{L}) \in
\overline{\mathcal{A}_{n,\hel,1}}$
and
$\langle \mathcal{L}\,a\,a{+}1\rangle = 0$ for some $a$, then
$\langle \mathcal{L}\,a\,a{+}2\rangle$ must necessarily be
non-positive\footnote{\label{ignoretwisted}
Unless $a \in \{n-1, n\}$, when one
must take into account the twisted cyclic symmetry.
In all that follows we will for simplicity always assume
that indices are outside of this range, which lets us
uniformly ignore all sign factors that
might arise from the twisted cyclic symmetry; these
signs necessarily
always conspire to ensure that all statements about
amplitudes are $\mathbb{Z}_n$ cyclically invariant.};
the proof of this assertion, which we omit here, parallels that of
a closely related statement proven in Sec.~6
of~\cite{Arkani-Hamed:2017vfh}.

Consider now a line of the form
$\mathcal{L} = (\alpha Z_a + \beta Z_{a+1}, A)$ for some point $A$
and some parameters
$\alpha, \beta$ which are not both vanishing.
We will show that an $\mathcal{L}$ of this form can lie in the
closure of an amplituhedron only if
$\mathcal{L} = (a\,a{+}1)$ or
$\alpha \beta \ge 0$.

First, as just noted, since
$\langle \mathcal{L}\,a\,a{+}1\rangle = 0$ we must have
\begin{align}
\label{eqn:a}
0 \ge \langle \mathcal{L}\,a\,a{+}2\rangle
= \beta \langle a{+}1\,A\,a\,a{+}2\rangle\,.
\end{align}
On the other hand, as mentioned at the end of~\secRef{amplituhedron},
we also have
$\langle \mathcal{L}\,a\,a{+}1 \rangle \ge 0$ for all $a$.
Applying this to $a+1$ gives
\begin{align}
\label{eqn:b}
0 \le  \langle \mathcal{L}\,a{+}1\,a{+}2\rangle =
\alpha \langle a\,A\,a{+}1\,a{+}2\rangle\,.
\end{align}
If $\langle a\,a{+}1\,a{+}2\,A\rangle \ne 0$, then the
two inequalities~(\ref{eqn:a}) and~(\ref{eqn:b}) imply that
$\alpha \beta \ge 0$.

This is the conclusion we wanted, but it remains to address
what happens if $\langle a\,a{+}1\,a{+}2\,A\rangle = 0$.
In this case $\mathcal{L}$ lies in the plane $(a\,a{+}1\,a{+}2)$
so we can take $\mathcal{L} =
(\alpha Z_a + \beta Z_{a+1}, \gamma Z_{a+1} + \delta Z_{a+2})$.
Then we have
\begin{align}
\begin{aligned}
0 &\ge \langle \mathcal{L}\, a{+}1\, a{+}3\rangle
= - \alpha \delta \langle a\, a{+}1\, a{+}2\, a{+}3\rangle\,,
\cr
0 &\le \langle \mathcal{L}\,a{-}1\,a\rangle =
\beta \delta \langle a{-}1\,a\,a{+}1\,a{+}2\rangle\,.
\end{aligned}
\end{align}
Both of the four-brackets in these inequalities are positive (for
generic projected external data) since
they are of the form
$\langle a\,a{+}1\,b\,b{+}1\rangle$, so we conclude that either
$\delta = 0$, which means that $\mathcal{L} = (a\,a{+}1)$,
or else we again have $\alpha \beta \ge 0$.

In conclusion, we have developed a robust test which establishes that
\begin{align}
\mathcal{L} = (\alpha Z_a + \beta Z_{a+1}, A) \in
\overline{\mathcal{A}_{n,\hel,1}}
\mbox{ only if }
\mathcal{L} = (a\,a{+}1)
 \mbox{ or }
\alpha \beta \ge 0\,.
\label{eqn:criterion}
\end{align}
This statement is independent of $\hel$ (and $Y$), but when
applied to particular branches, we will generally
encounter cases for which $\alpha \beta$ is negative unless
certain sequences of
four-brackets of the projected external data have a certain number
of sign flips; this signals that the branch
may intersect $\overline{\mathcal{A}_{n,\hel,1}}$ only
for certain values of $\hel$.

\subsection{MHV Lower Bounds}
\label{sec:mhvlower}

The fact that MHV amplituhedra only have boundaries of type
(1)--(7), (10) and (12)
(referring to the numbers given in the ``Branches'' column
of~\tabRef{bigtable})
follows implicitly from
the results of~\cite{Dennen:2016mdk} where all boundaries
of one- (and two-) loop MHV amplituhedra were studied.
It is nevertheless useful to still consider these cases
since we will need the corresponding $D$-matrices below to
establish that amplituhedra have boundaries of these types
for all $0 \le \hel \le n-4$.

In this and the following two sections we always assume, without
loss of generality, that indices $i, j, k, \ell$ are cyclically
ordered and non-adjacent ($i{+}1 < j < j{+}1 < k< k{+}1 < \ell$), and moreover
that $1 < i$ and $\ell < n$.  In particular, this means that
we ignore potential signs from the twisted
cyclic symmetry (see footnote~\ref{ignoretwisted}).

\paragraph{Branch (4)} is a prototype
for several other branches, so we begin with it
instead of branch (1).
The solution for $\mathcal{L}$ shown
in~\tabRef{bigtable} may be represented as $\mathcal{L} = D Z$ with
\begin{align}
\label{eqn:twocutdmatrix}
D =
\bordermatrix{ & i & i{+}1&  j & j{+}1 \cr
 &\alpha &1- \alpha & 0 &0 \cr
 & 0  & 0 & \beta &1- \beta}\,,
\end{align}
where we display only the nonzero columns of the $2 \times n$ matrix
in the indicated positions $i$, $i{+}1$, $j$ and $j{+}1$.
This solves the two-mass bubble on-shell
conditions for all values of the parameters $\alpha$ and $\beta$.
This branch intersects $\overline{\mathcal{A}_{n,0,1}}$
when they lie in the range $0 \le \alpha, \beta \le 1$, where
the matrix $D$ is non-negative.
Thus we conclude that MHV amplituhedra have boundaries of type (4).

\paragraph{Branches (5), (6), (7), (10), and (12)} can all
be represented by special cases of~\eqnRef{twocutdmatrix}
for $\alpha$ and/or $\beta$ taking values 0 and/or 1,
and/or with columns relabeled, so
MHV amplituhedra also have boundaries of all of these types.

\paragraph{Branch (1)} may be represented by
\begin{align}
\label{eqn:tadpole}
D =
\bordermatrix{ & & i{-}1 & i & i{+}1 & i{+}2  \cr
&\cdots & 0 & \alpha & 1-\alpha & 0 & \cdots \cr
& \cdots& \alpha_{i-1} & \alpha_{i} & \alpha_{i+1} & \alpha_{i+2} &\cdots }\,.
\end{align}
This provides a solution to the tadpole on-shell condition
$\langle \mathcal{L}\,i\,i{+}1\rangle = 0$
for all values of the parameters,
and there clearly are ranges for which $D$ is non-negative.
Note that all but two of the parameters in the second row could be gauged
away, but this fact is not relevant at the moment (see
footnote~\ref{footnote10}).
If $0 \le \alpha \le 1$, we could have either
$\alpha_a = 0$ for $a < i+1$ and $\alpha_a > 0$ for $a>i$, or
$\alpha=0$ for $a > i$ and $\alpha_a < 0$ for $a < i+1$.
We conclude that MHV amplituhedra also have boundaries of this
type.

\paragraph{Branch (2)} is the special case $\alpha=1$ of branch (1).

\paragraph{Branch (3)} may be represented by
\begin{align}
D = \bordermatrix{  & i{-}1 & i & i{+}1   \cr
& 1 & 0 & \alpha \cr
 & 0  & 1 & \beta }
\end{align}
for arbitrary $\alpha, \beta$, which is non-negative
for $\alpha \le 0$ and $\beta \ge 0$, so
MHV amplituhedra also have boundaries of this
type.

\subsection{NMHV Lower Bounds}
\label{sec:nmhvlower}

\paragraph{Branch (8)}
of the two-mass triangle may be represented as
\begin{align}
\label{eqn:two-mass-triangle-d}
D=
\bordermatrix{ &  i & i{+}1 & j & j{+}1 \cr
  & \alpha & 1-\alpha  & 0 &  0 \cr
  & 0 & 0  & -\langle \overline{i} \, j{+}1\rangle & \langle \overline{i}\, j \rangle}
\end{align}
for arbitrary $\alpha$.
For generic projected external data $\mathcal{L} \ne
(j\,j{+}1)$, so criterion~(\ref{eqn:criterion}) shows
that this configuration has a chance to lie on the boundary of an
amplituhedron only if $-\langle \overline{i}\, j{+}1\rangle
\langle \overline{i}\,j\rangle \ge 0$.
This is not possible for MHV external data,
where the ordered four-brackets are always positive,
so MHV amplituhedra do not have boundaries of this type.
But note that the inequality can be satisfied
if there is at least one sign flip in the
sequence $\langle \overline{i}\, \bullet \rangle$,
between $\bullet = j$ and $\bullet = j{+}1$.
This motivates us to consider $\hel = 1$, so let us now check that with
\begin{align}
\label{eqn:branch8c}
C =
\bordermatrix{
& {i{-}1} & i & {i{+}1} & j & {j{+}1} \cr
& c_{i-1} & c_i & c_{i+1} & c_j & c_{j+1}
}\,,
\end{align}
the pair $C, D$ is a valid configuration.
First of all, it is straightforward to check that $\mathcal{L}=D
\mathcal{Z}$ still satisfies the two-mass triangle on-shell
conditions.  This statement is not completely trivial since these
conditions now depend on $Y = C \mathcal{Z}$ because
of the projection~(\ref{eqn:projection}).
Second, in order for $C$ to be non-negative we need all five of the
indicated $c_a$'s to be non-negative.
Moreover, in order to support generic projected external data, we need
them all to be nonzero --- if, say, $c_i$ were equal to zero, then
$\langle i{-}1\,i{+}1\,j\,j{+}1\rangle$ would vanish, etc.
Finally, for
$\left( \begin{smallmatrix} D \\ C \end{smallmatrix} \right)$
to be non-negative we need
\begin{align}
0 \le \alpha \le \frac{c_i}{c_i+ c_{i+1}}\,.
\end{align}
This branch intersects $\overline{\mathcal{A}_{n,1,1}}$
for $\alpha$ in this range, so we conclude that NMHV
amplituhedra have boundaries of this type.

\paragraph{Branch (9)}
is the general solution of the three-mass triangle, and is
already given in~\tabRef{bigtable} in $D$-matrix form as
\begin{align}
D =
\bordermatrix{ & i & i{+}1  & j & j{+}1 \cr
& \alpha & 1-\alpha & 0 & 0 \cr
& 0 & 0   & \rho(\alpha) & \sigma(\alpha)  }\,,
\end{align}
with $\rho(\alpha)$ and $\sigma(\alpha)$ defined in~\eqnRef{rhosigma}.
For generic projected external data this $\mathcal{L}$ can never
attain the value $(i\,i{+}1)$ or $(j\,j{+}1)$.  Applying
criterion~(\ref{eqn:criterion}) for both $a=i$ and $a=j$ shows
that this configuration has a chance to lie on the boundary of an
amplituhedron only if
$\alpha(1-\alpha) \ge 0$ and $\rho(\alpha) \sigma(\alpha) \ge 0$.
This is not possible for MHV external data, so we conclude
that MHV amplituhedra do not have boundaries of this type.
However, the $\rho(\alpha) \sigma(\alpha) \ge 0$ inequality can be satisfied
if the sequences $\langle i\, k\, k{+}1\, \bullet\rangle$ and
$\langle i{+}1\,k\,k{+}1\,\bullet\rangle$ change sign
between $\bullet = j$ and $\bullet = j{+}1$, as long
as the sequences $\langle j\,k\,k{+}1\,\bullet\rangle$ and
$\langle j{+}1\,k\,k{+}1\,\bullet\rangle$ do not flip sign here.
Consider for $\hel=1$ the matrix
\begin{align}
\label{eq:triangleC}
C =
\bordermatrix{
& i & {i{+}1} & j & {j{+}1} & k & {k{+}1} \cr
& \alpha c_i & (1-\alpha)c_i & c_j & c_{j+1} & c_k & c_{k+1}
}\,.
\end{align}
Then $C, D$ is a valid configuration because
(1) $\mathcal{L} = D \mathcal{Z}$
satisfies the three-mass triangle on-shell conditions
(for all values of $\alpha$ and the $c$'s),
and, (2) for $0 \le \alpha \le 1$
and all $c$'s positive,
the $C$-matrix is non-negative and supports generic
positive external data (because it
has at least $\hel{+}4=5$ nonzero columns), and
(3) for this range of
parameters $\left( \begin{smallmatrix} D \\ C \end{smallmatrix} \right)$
is also non-negative.
Since this branch intersects
$\overline{\mathcal{A}_{n,1,1}}$ for a range of $\alpha$,
we conclude that NMHV amplituhedra have boundaries of this type.

\paragraph{Branch (16)} is the special case $\alpha =1$ of
branch (9).

\paragraph{Branch (14)}  is the special case
$j \to i+1$, $k \to j$ of branch (16).

\paragraph{Branch (15)}  is equivalent to the mirror image
of branch (14), after relabeling.

\paragraph{Branch (11)} is the special case $j = i+2$ of
branch (15).

\subsection{N\texorpdfstring{${}^2$}{N}MHV Lower Bounds}
\label{sec:nnmhvlower}

\paragraph{Branch (17)} may be represented by
\begin{align}
\label{eqn:three-mass-box}
D=
\bordermatrix{ &j & j{+}1 & k & k{+}1 \cr
& 0 & 0 & -\langle \overline{i}\, k{+}1 \rangle & \langle \overline{i}\, k \rangle \cr
 & -\langle \overline{i}\, j{+}1 \rangle  & \langle \overline{i}\, j \rangle &  0 & 0}\,.
 \end{align}
For generic projected external data the
corresponding $\mathcal{L}$ will never attain the value $(j\,j{+}1)$ or
$(k\,k{+}1)$.
We can apply criterion~(\ref{eqn:criterion}) for both $a=j$ and $a=k$,
which reveals that this configuration has a chance to lie
on a boundary of an amplituhedron only if
both $-\langle \overline{i}\,j{+}1\rangle \langle \overline{i}\,j\rangle
\ge 0$ and
$-\langle \overline{i}\,k{+}1\rangle \langle \overline{i}\,k\rangle \ge 0$.
This is impossible for MHV external data, and it is also impossible
in the NMHV case,
where some projected four-brackets may be negative but the sequence
$\langle \overline{i}\,\bullet\rangle$ may only flip sign once,
whereas we need it to flip sign twice,
once between $\bullet = j$ and $\bullet =j{+}1$, and again
between $\bullet = k$ and $\bullet = k{+}1$.
We conclude that $\hel < 2$ amplituhedra do not have
boundaries of this form.
Consider now pairing~(\ref{eqn:three-mass-box}) with the $\hel = 2$
matrix
\begin{align}
C =
\bordermatrix{ &i{-}1 &i &i{+}1&j & j{+}1 & k & k{+}1 \cr
&c_{11} &c_{12} & c_{13} & c_{14} & c_{15} &0 & 0 \cr
&c_{21} &c_{22} & c_{23} &0 &0 & c_{24} & c_{25} \cr } \,.
\label{eqn:threemassc}
\end{align}
It is straightforward to check that
$C, D$ is a valid configuration for a range of values of $c$'s, so we conclude
that $\hel = 2$ amplituhedra have boundaries of this type.

\paragraph{Branch (13)} may be represented by
\begin{align}
D=
\bordermatrix{ & i{-}1& i & i{+}1 \cr
 & \langle i \, \overline{j} \rangle & -\langle i{-}1 \, \overline{j} \rangle &  0 \cr
 & 0  & - \langle i{+}1 \, \overline{j} \rangle &  \langle i \, \overline{j} \rangle}\,,
\end{align}
which by~(\ref{eqn:criterion}) cannot lie on a boundary of an amplituhedron
unless the sequence $\langle \overline{j}\, \bullet\rangle$
flips sign twice,
first between $\bullet = i{-}1$ and $i$
and again between $\bullet = i$ and $i{+}1$.
Therefore, neither MHV nor NMHV amplituhedra have boundaries of this type.
However it is straightforward to verify that with
\begin{align}
C = \bordermatrix{ &i{-}1 &i &i{+}1 & j{-}1 & j & j{+}1 \cr
&c_{11} &c_{12}&0 & c_{13} & c_{14} & c_{15} \cr
&0 &c_{21} &c_{22} & c_{23} & c_{24} & c_{25} }
\end{align}
the pair $C, D$ is a valid configuration for a range of values of $c$'s,
so $\hel = 2$ amplituhedra do have boundaries of this type.

\paragraph{Branches (18) and (19)}
of the four-mass box may be represented as
\begin{align}
D =
\bordermatrix{ & i & i{+}1 & j & j{+}1 \cr
 & \alpha & 1-\alpha & 0 & 0 \cr
 & 0  & 0 & \beta & 1-\beta }\,,
 \end{align}
where $\alpha$ and $\beta$ are fixed by requiring that $\mathcal{L}$
intersects the lines $(k \, k{+}1)$ and $(\ell \, \ell{+}1)$.
The values of $\alpha$ and $\beta$ on the two branches
were written explicitly in~\cite{Bourjaily:2013mma};
however, the complexity of those expressions makes analytic
positivity analysis difficult.
We have therefore resorted to numerical testing:
using the algorithm described in Sec.~5.4 of~\cite{ArkaniHamed:2012nw}, we
generate a random positive $n \times (\hel+4)$
$\mathcal{Z}$-matrix and
a random positive $\hel \times n$ $C$-matrix.
After projecting through
$Y = C \mathcal{Z}$, we obtain projected external data with the correct
$\nkmhv{\hel}$ sign-flipping properties.
We have checked numerically
that both four-mass box branches lie on the boundary
of $\nkmhv{\hel}$ amplituhedra only for $\hel \ge 2$,
for many instances of randomly generated external data.

\subsection{Emergent Positivity}
\label{sec:emergent}

The analysis of Secs.~\ref{sec:mhvlower}, \ref{sec:nmhvlower}
and~\ref{sec:nnmhvlower} concludes the proof
of all of the lower bounds on $\hel$
shown in the fourth column of~\tabRef{bigtable}.
We certainly do not claim to have written down the most
general possible valid $C, D$ configurations; the ones we display
for $\hel > 0$
have been specifically chosen to demonstrate an
interesting feature we call \emph{emergent positivity}.

In each $\hel > 0$ case we encountered $D$-matrices
that are only non-negative if certain sequences
of projected four-brackets of the form $\langle a\,a{+}1\,b\,\bullet\rangle$
change sign $\hel$ times, at certain precisely
specified locations.
It is straightforward to check that within
the range of validity of each $C, D$ pair we have
written down, the structure of the $C$ matrix is such that
it automatically puts
the required sign flips in just the right places to make the $D$
matrix, on its own, non-negative
(provided, of course, that
$\left( \begin{smallmatrix} D \\ C \end{smallmatrix} \right)$
is non-negative).
It is not a priori obvious that it had to be possible to find pairs $C, D$
satisfying this kind of emergent positivity;
indeed, it is easy to find valid pairs
for which it does not hold.

\subsection{Parity and Upper Bounds}
\label{sec:parity}

Parity relates each branch to itself or to the other branch associated
with the same Landau diagram.
Since parity is a symmetry of the amplituhedron~\cite{Arkani-Hamed:2017vfh}
which relates $\hel$ to $n - \hel - 4$,
the lower bounds on $\hel$ that we have established
for various branches imply upper bounds on $\hel$ for their
corresponding parity conjugates.
These results are indicated in the fourth column
of~\tabRef{bigtable}, where the inequalities are aligned
so as to highlight the parity symmetry.

Although these $\hel$ upper bounds are required by parity
symmetry, they may seem rather mysterious
from the analysis carried out so far.
We have seen that certain branches can be boundaries
of an amplituhedron only if certain sequences of four-brackets
have (at least) one or two sign flips.
In the next section, we explain a mechanism
which gives an upper bound to the number of sign flips,
or equivalently which gives  the upper bounds on
$\hel$ that are required by parity symmetry.

\subsection{Increasing Helicity}
\label{sec:largek}

So far we have only established that
$\nkmhv{\hel}$ amplituhedra have boundaries
of certain types for specific low (or, by parity symmetry, high)
values of $\hel$.  It remains to show that all of the branches
listed in~\tabRef{bigtable} lie on boundaries of amplituhedra
for all of the intermediate helicities.
To this end we describe now an algorithm for converting
a valid configuration $C_{0}, D_{0}$ at the initial, minimal
value of $\hel_0$ (with $C_{0}$ being
the empty matrix for those branches with $\hel_0 = 0$)
into a configuration that is valid at some higher value of $\hel$.

We maintain the structure of $D \equiv D_{0}$
and append to $C_0$ a matrix $C'$ of dimensions
$(\hel - \hel_0) \times n$
in order to build a configuration for helicity $\hel$.
Defining $C=\left( \begin{smallmatrix} C_0 \\ C' \end{smallmatrix} \right)$,
we look for a $C'$ such that following properties are satisfied:
\begin{enumerate}
\item The same on-shell conditions are satisfied.
\item
In order for the configuration to support generic projected
external data, the $C$-matrix must have $m \ge \hel + 4$ nonzero columns,
and the rank of any $m-4$ of those columns must be $\hel$.
\item Both $C$ and
$\left( \begin{smallmatrix} D \\ C \end{smallmatrix} \right)$
remain non-negative.
\end{enumerate}	
Since the $C$-matrix only has $n$ columns in total,
it is manifest from property (2) that everything shuts off for
$\hel > n -4$, as expected.

Let us attempt to preserve the emergent positivity of $D$.
If $\hel_0 = 0$ then this
is trivial;
the $D$-matrices in~\secRef{mhvlower} do not depend on
any brackets, so adding rows to the empty $C_0$ has no effect on $D$.
For $\hel_0>0$, let $A$ and $B$ be two
entries in $D_0$ that are responsible for imposing a sign
flip requirement.
The argument applies equally to all of the $\hel_0 > 0$
branches, but for the sake
of definiteness consider
from~\eqnRef{two-mass-triangle-d}
the two four-bracket dependent entries
$A = - \langle \overline{i}\,j{+}1\rangle$ and
$B = \langle \overline{i}\,j\rangle$.
Assuming that $C_0$ is given by~\eqnRef{branch8c}
so that both $A$ and $B$ are positive with respect to
$Y_0 = C_0 \mathcal{Z}$, then
$A B = - [Y_0 \, \overline{i} \, j{+}1] [Y_0 \, \overline{i} \, j] > 0$.
If we append a second row $C'$
and define $Y' = C' \mathcal{Z}$
then we have
\begin{equation}
\begin{aligned}
A &
=
- [ Y_0\, Y'\,\mathcal{Z}_{i-1}\,\mathcal{Z}_i\,\mathcal{Z}_{i+1}\, \mathcal{Z}_{j+1} ]
=
- c_j [ \mathcal{Z}_j\, Y'\,\mathcal{Z}_i\,\mathcal{Z}_{i+1}\,\mathcal{Z}_{j+1} ]
\,,
\cr
B &
=
[ Y_0\,Y'\,\mathcal{Z}_{i-1}\,\mathcal{Z}_i\,\mathcal{Z}_{i+1}\,\mathcal{Z}_j ]
=
c_{j+1} [ \mathcal{Z}_{j+1}\,Y'\,\mathcal{Z}_i\,\mathcal{Z}_{i+1}\,\mathcal{Z}_{j}]\,.
\end{aligned}
\end{equation}
Since $c_{j}$ and $c_{j+1}$ are both positive, we see that $A$ and $B$
still satisfy $A B > 0$, regardless of the value of $Y'$.
By the same argument,
arbitrary rows can be added to a $C$-matrix without affecting
the on-shell conditions, so property (1) also holds trivially (and
also if $\hel_0 = 0$).

The structure of the initial
$D_0$
of Secs.~\ref{sec:mhvlower}, \ref{sec:nmhvlower}
and~\ref{sec:nnmhvlower} are similar in that the nonzero
columns of this matrix are grouped into at most two \emph{clusters}%
\footnote{Branch (1)
appears to be an exception, but only because~\eqnRef{tadpole}
\label{footnote10}
as written
is unnecessarily general: it is sufficient for the second row to have only
three nonzero entries, either in columns $\{i{-}3,i{-}2,i{-}1\}$
or in columns $\{i{+}1,i{+}2,i{+}3\}$.}.
For example, for branch (17) there are two clusters
$\{j,j{+}1\}$ and $\{k,k{+}1\}$ while
for branch (3) there is only a single cluster
$\{i{-}1,i,i{+}1\}$.
Property (3) can be preserved most easily
if we add suitable columns only in
a \emph{gap} between clusters.
Let us illustrate how this works in the case of branch (4)
where $C_0$ is empty and we can start by taking either
\begin{align}
\label{eqn:dccut1}
\begin{pmatrix}
D_0 \cr
C\cr
\end{pmatrix}
=
\bordermatrix{
& &i{-}1 & i & i{+}1&i{+}2&\cdots &j{-}1 &  j & j{+}1 &j{+}2 & \cr
 &\cdots & 0 &\alpha & 1-\alpha & 0&\cdots & 0&0 &0 &0 & \cdots \cr
&\cdots & 0 &0 & 0&0 &\cdots&  0&\beta &1-\beta &0 & \cdots \cr
&\cdots & 0 &0&\vec{c}_{i+1} &\vec{c}_{i+2}  &\cdots & \vec{c}_{j-1}& \vec{c}_j& 0 & 0 & \cdots \cr
 }
\end{align}
to fill in the gap between clusters $\{i, i{+}1\}$
and $\{j, j{+}1\}$, or
\begin{align}
\label{eqn:dccut2}
\begin{pmatrix}
D_0 \cr
C\cr
\end{pmatrix}
=
\bordermatrix{
& &i{-}1 & i & i{+}1&i{+}2&\cdots &j{-}1 &  j & j{+}1 &j{+}2 & \cr
 &\cdots & 0 &\alpha & 1-\alpha & 0&\cdots & 0&0 &0 &0 & \cdots \cr
&\cdots & 0 &0 & 0&0 &\cdots&  0&\beta &1-\beta &0 & \cdots \cr
&\cdots & \vec{c}_{i-1} &\vec{c}_i&0 &0  &\cdots & 0& 0& \vec{c}_{j+1} & \vec{c}_{j+2} & \cdots \cr
}
\end{align}
to fill in the gap between $\{j, j{+}1\}$ and $\{i, i{+}1\}$ that
``wraps around'' from $n$ back to~$1$.
In both~(\ref{eqn:dccut1}) and~(\ref{eqn:dccut2}) each $\vec{c}_a$
is understood to be a $\hel$-component column vector, and in both cases
$\left( \begin{smallmatrix} D_0 \\ C \end{smallmatrix} \right)$
can be made non-negative
as long as $C$ is chosen to be non-negative\footnote{If $\hel$
is even this is automatic; if $\hel$ is odd the
two rows of $D_0$ should be exchanged.}.
In this manner we can trivially increment
the $\hel$-validity of a given configuration until the gaps become full.
This cutoff depends on the precise positions of the gaps,
and is most stringent when the two clusters are maximally separated
from each other,
since this forces the gaps to be relatively small.
In this worst case
we can fit only $\lceil\frac{n}{2}\rceil$
columns into a $C$-matrix of one of the above two types.
Keeping in mind property (2) that the $C$-matrix should have
at least $\hel + 4$ nonzero columns,
we see that this construction can reach values of $\hel
\le \lceil{\frac{n}{2}}\rceil-4$.
In order to proceed further, we can (for example)
add additional columns $c_i$ and $c_{j+1}$
to~\eqnRef{dccut1}, or $c_{i+1}$ and $c_j$
to~\eqnRef{dccut2}.  Choosing
a non-negative $C$ then no longer trivially
guarantees
that $\left( \begin{smallmatrix} D_0 \\ C \end{smallmatrix} \right)$
will also be non-negative, but there are ranges of $C$ for which this
is possible to arrange, which is sufficient for our argument.

It is possible to proceed even further by adding additional,
specially crafted columns in both gaps,
but the argument is intricate and depends delicately
on the particular
structure of each individual branch (as evident from the
delicate structure of $\hel$ upper
bounds in~\tabRef{bigtable}).
In the interest of brevity we terminate our discussion
of the algorithm here and
note that it is straightforward to check that for all boundaries,
even in the worst case the gaps are always big enough
to allow the construction we have described to proceed up to
and including the parity-symmetric
midpoint $\hel = \lfloor{\frac{n}{2}}\rfloor-2$; then
we appeal again to parity symmetry in order to establish the existence
of valid configurations for $\hel$ between this midpoint and the upper bound.

\bigskip
\noindent
This finally concludes the proof of the $\hel$-bounds
shown in the fourth column of~\tabRef{bigtable}, and
thereby step 1(c) from~\secRef{summary}.

\section{The Hierarchy of One-Loop Boundaries}
\label{sec:hierarchy}

Step (1) of our analysis (\secRef{summary}) is now complete at one loop.
Before moving on to step (2) we demonstrate
that the boundaries classified in~\secRef{one-loop-boundaries}
can be generated by a few simple graph operations
applied to the maximal codimension boundaries of MHV amplituhedra
(\tabRef{bigtable} type (12) or, as a special case, (10)).
This arrangement will prove useful in the sequel since
one-loop boundaries are the basic building blocks for
constructing boundary configurations at arbitrary loop order.

We call boundaries of
type (2), (5)--(7), (10), (12), and (14)--(16)
\emph{low-$\hel$} boundaries
since they are valid for the smallest value of $\hel$ for their
respective Landau diagrams.
The branches (8), (11), (13) and (17) are
\emph{high-$\hel$} boundaries and
are respectively the parity conjugates
of (7), (10), (12) and (16).
Branch (3), the parity conjugate of branch (2), is properly regarded
as a high-$\hel$ boundary since (2) is low-$\hel$,
but it is accidentally valid for all $\hel$.
Branches (1), (4), and (9) are self-conjugate
under parity and are considered both low-$\hel$ and
high-$\hel$, as are the parity-conjugate pair
(18), (19).

\subsection{A Graphical Notation for Low-helicity Boundaries}
\label{sec:graphical-notation}

\begin{figure}
\centering
\begin{tabular}
{
>{\centering\arraybackslash} m{0.25\textwidth} 
>{\centering\arraybackslash} m{0.25\textwidth} 
>{\centering\arraybackslash} m{0.25\textwidth}  
}
\includegraphics{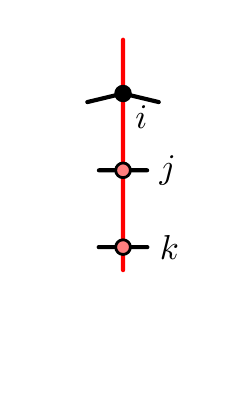}
&
\includegraphics{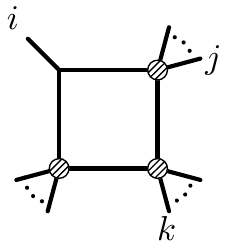}
&
\includegraphics{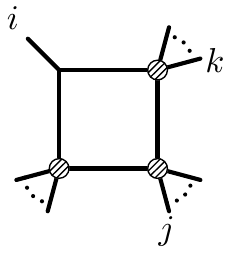}
\\[-20pt]
  (a) & (b) & (c)
              \\
\end{tabular}
\caption[Explicit twistor solution diagrams.]{%
The twistor diagram shown in (a) depicts
branch (16) of solutions to the
three-mass box
on-shell conditions
$\langle \mathcal{L}\,i{-}1\,i\rangle =
\langle \mathcal{L}\,i\,i{+}1\rangle =
\langle \mathcal{L}\,j\,j{+}1\rangle =
\langle \mathcal{L}\,k\,k{+}1\rangle = 0$, which is a valid
boundary for $\hel \ge 1$.
This branch passes through the point $Z_i$ and intersects
the lines $(j\,j{+}1)$ and $(k\,k{+}1)$.
As drawn, the intersection at $j$ is an example
of a non-MHV intersection, but the figure is agnostic about the relative
cyclic ordering of $i, j, k$ and is intended to
represent either possibility.
Therefore, the corresponding Landau diagram can be either (b) or (c)
depending on whether
$i < j < k$ or $i < k < j$.
}
 \label{fig:explicit-graphical-notation}
\end{figure}

We begin by devising a graphical notation
in terms of which the operations between
momentum twistor solutions are naturally phrased.
These graphs are \emph{twistor diagrams}\footnote{Not
to be confused with the twistor diagrams
of~\cite{Hodges:2005bf}.}
depicting various configurations of intersecting lines
in $\mathbb{P}^3$.
The elements of a twistor diagram, an example of
which is shown in panel (a) of~\figRef{explicit-graphical-notation},
are:
\begin{itemize}
\item The red line depicts an
$\mathcal{L}$ solving some on-shell conditions, specifically:
\item if $\mathcal{L}$ and a single line segment
labeled $i$ intersect at an empty node,
then \\
$\langle \mathcal{L} \, i \, i{+}1 \rangle =0$, and
\item
if $\mathcal{L}$ and two line segments intersect at a filled node labeled $i$, then \\
$\langle \mathcal{L} \, i{-}1 \, i \rangle = \langle \mathcal{L} \, i \, i{+}1 \rangle =0$.
\end{itemize}
An ``empty'' node is colored red, indicating the line passing through it.
A ``filled'' node is filled in solid black, obscuring the line passing through it.

In general a given $\mathcal{L}$ can pass through as many as
four labeled nodes (for generic projected external data, which
we always assume).
If there are four, then none of them can be filled.  If there
are three, then at most one of them can be filled, and we choose
to always draw it as either the first or last node along $\mathcal{L}$.
If there are more than two, then any nodes between
the first and last are called \emph{non-MHV intersections}, which
are necessarily empty.
This name is appropriate because
branches satisfying such on-shell constraints
are not valid boundaries
of MHV amplituhedra, and each non-MHV intersection in a twistor
diagram increases the minimum value of $\hel$ by one.

%
%
%
%
%
%

Although no such diagrams
appear in this paper, the extension to higher
loops is obvious:  each $\mathcal{L}$ is represented by a line of a different
color, and
the presence of an on-shell condition of the form
$\langle \mathcal{L}^{(\ell_1)} \, \mathcal{L}^{(\ell_2)}\rangle = 0$
is indicated by an empty node at the intersection of the lines
$\mathcal{L}^{(\ell_1)}$ and
$\mathcal{L}^{(\ell_2)}$.

To each twistor diagram it is simple to associate one or more
Landau
diagrams, as also shown
in~\figRef{explicit-graphical-notation}.
If a twistor diagram has a filled node at $i$ then an
associated Landau diagram has two propagators
$\langle \mathcal{L}\,i{-}1\,i\rangle$ and $\langle
\mathcal{L}\,i\,i{+}1\rangle$
requiring a massless corner at $i$ in the Landau diagram.
If a twistor diagram has an empty node on the line segment
marked $i$ then an associated Landau diagram
only has the single propagator
$\langle \mathcal{L}\,i\,i{+}1\rangle$,
requiring a massive corner in the Landau diagram.
Therefore, twistor diagrams should be thought of as graphical shorthand
which both depict the low-$\hel$ solution to the cut conditions and
simultaneously represent one or more Landau diagrams, as explained
in the caption of~\figRef{explicit-graphical-notation}.

One useful feature of this graphical
notation is that the nodes of a twistor diagram
fully encode the total number of propagators,
$\nProps$, in the Landau diagram (and so also the total number
of on-shell conditions):
each filled node accounts for two propagators,
and each empty node accounts for one propagator:
\begin{align}
\nProps = 2 \nFull + \nEmpty\,.
\label{eqn:propfullemptyrelation}
\end{align}
This feature holds at higher loop order where this counting
directly indicates how many propagators to associate with each loop.

Let us emphasize that a twistor diagram generally contains more information
than its associated Landau diagram, as it indicates not only the set of
on-shell conditions satisfied, but also specifies a particular branch
of solutions thereto.
The sole exception is the four-mass box,
for which the above rules do not provide the twistor diagram
with any way to distinguish the two branches (18), (19) of solutions.
Moreover,
the rules also do not provide any way to indicate that an $\mathcal{L}$
lies in a particular plane, such as $\overline{i}$. Therefore
we can only meaningfully represent the low-$\hel$ boundaries
defined at the beginning of~\secRef{hierarchy}.

Given a twistor diagram depicting some branch, a twistor diagram
corresponding to a relaxation of that branch may be obtained
by deleting a non-MHV intersection of the type shown
in (a) of~\figRef{explicit-graphical-notation},
by replacing a filled node and its two line segments with an empty node
and a single segment, or by deleting an empty node.
In the associated Landau diagram, a relaxation corresponds to collapsing
an internal edge of the graph.
This is formalized in greater detail in~\secRef{graphical-organization}.

\subsection{A Graphical Recursion for Generating Low-helicity Boundaries}
\label{sec:graphical-organization}

%
%
\begin{figure}
\centering
\includegraphics{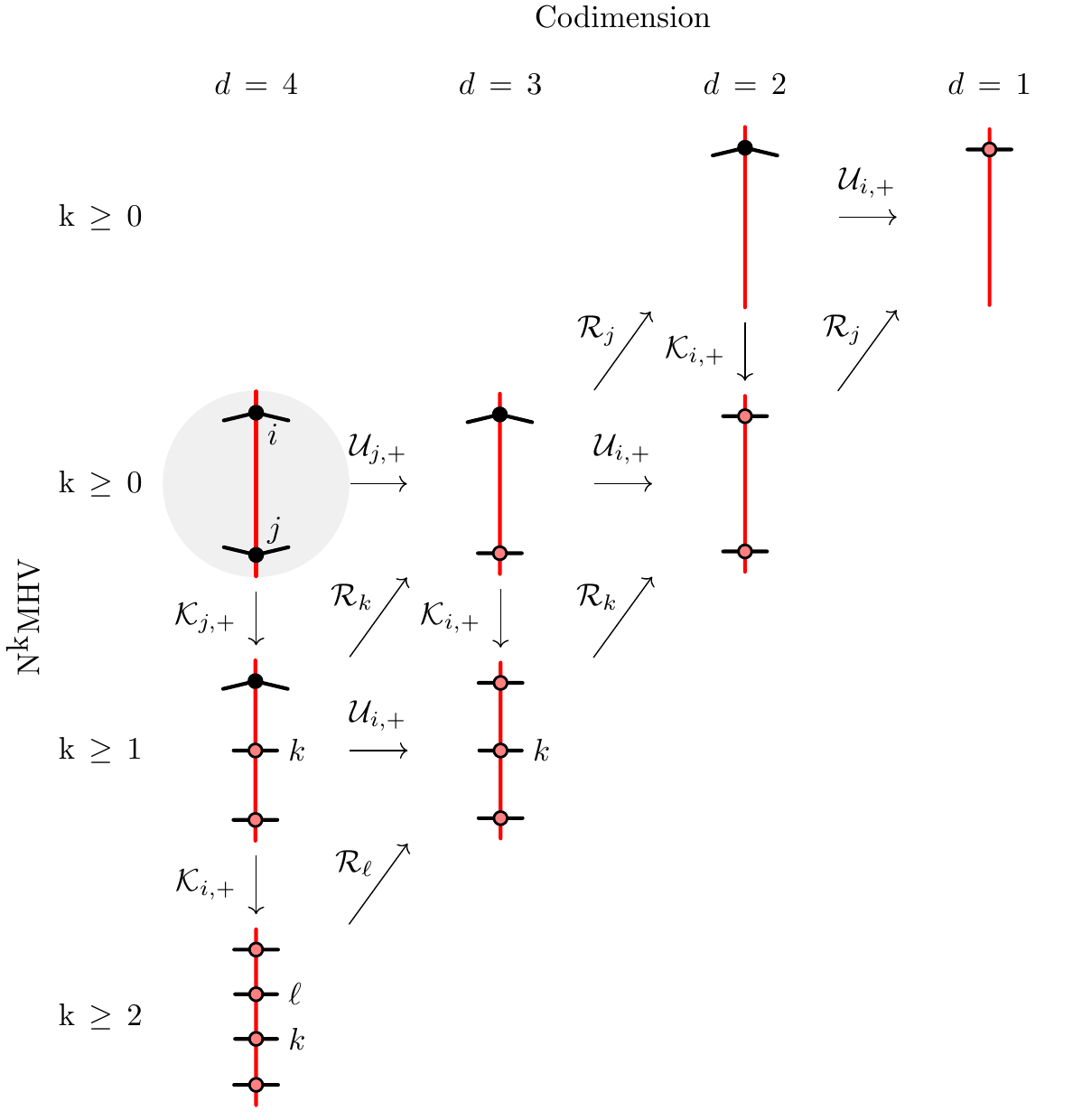}
\caption[One-loop maximal codimension graph flow.]{%
Twistor diagrams depicting eight types
of low-$\hel$ boundaries of $\nkmhv{\hel}$ amplituhedra,
organized according to the minimum value of $\hel$
and the codimension $d$ (equivalently, the number of
on-shell conditions satisfied).
These correspond respectively to branch
types (2), (1), (12), (7), (4), (16), (9) and (18)/(19).
The graph operators $\ko{}$, $\ro{}$, and $\uo{}$ are explained in the
text and demonstrated in~\figRefs{graph-operation}{unpin-go}, respectively.
Evidently all eight types of boundaries can be generated by acting
with sequences of these operators on
MHV maximal codimension boundaries of the type
shown shaded in gray.
There is an analogous parity-conjugated version of this hierarchy
which relates all of the high-$\hel$ branches to each other.
The missing low-$\hel$ boundary types (5), (6), (10), (11), (14) and
(15) are degenerate cases which can be obtained by starting with
$j=i+1$ in the gray blob.
}
 \label{fig:one-loop-graph-flow}
\end{figure}
%
%

In~\figRef{one-loop-graph-flow} we organize
twistor diagrams representing eight types of boundaries
according to $d$ and $\hel$; these are respectively
the number of on-shell conditions $d$ satisfied on the boundary,
and
the minimum value of $\hel$ for which the boundary is valid.
It is evident from this data
that there is a simple relation between
$d$, $\hel$, and the number of
filled ($\nFull$) and empty ($\nEmpty$) nodes.
Specifically, we see that
an $\nkmhv{\hel}$ amplituhedron can have boundaries of a type
displayed in a given twistor diagram only if
\begin{align}
\hel \ge  2\nEmpty + 3 \nFull - d - 2 = \nEmpty + \nFull - 2 \,,
\label{eqn:k-from-mts}
\end{align}
where we have used~\eqnRef{propfullemptyrelation} with
$\nProps = d$.
In the sequel we will describe a useful map from Landau
diagrams to the on-shell diagrams of~\cite{ArkaniHamed:2012nw}
which manifests the relation~(\ref{eqn:k-from-mts})
and provides a powerful generalization thereof to higher loop order.
The amplituhedron-based approach has some advantages over that
of enumerating on-shell diagrams that will also be explored
in the sequel.
First of all, the minimal required helicity of a multi-loop configuration
can be read off from each loop line separately.
Second, we immediately know the relevant solution branches for a given helicity.
And finally, compared to enumerating all relevant on-shell diagrams
the amplituhedron-based method is significantly more compact
since it can be used to produce a minimal subset of diagrams
such that all allowed diagrams are relaxations thereof,
including limits where massive external legs become
massless or vanish.

From the data displayed in~\figRef{one-loop-graph-flow}
we see that a natural organizational
principle emerges: all $\nkmhv{\hel}$ one-loop twistor diagrams can be obtained
from the unique maximal codimension MHV diagram (shown shaded in gray)
via sequences of
simple graph operations which we explain in turn.

%
%
\begin{figure}
\centering
\includegraphics{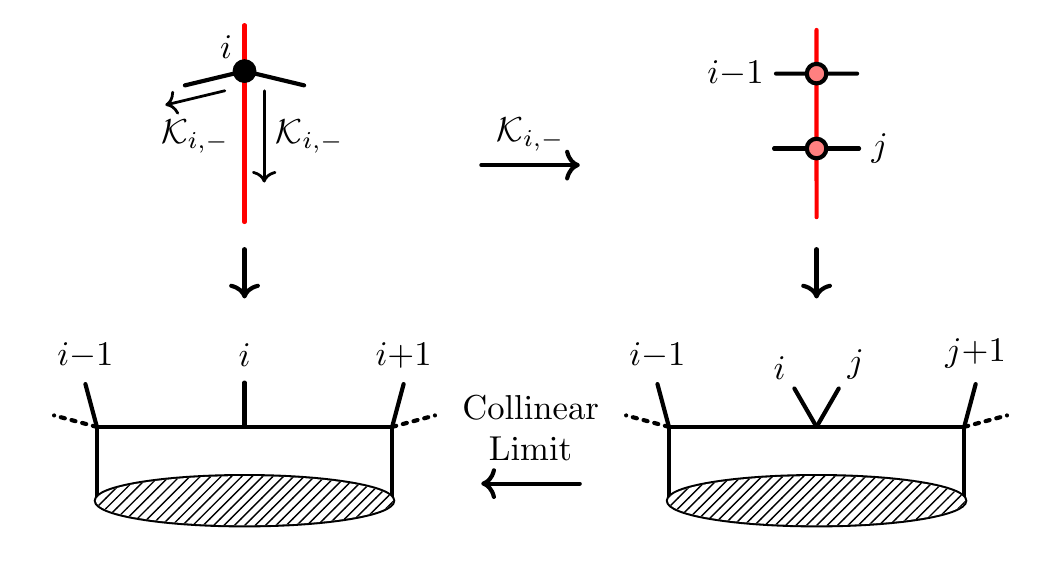}
\caption[Graph operation on momentum twistor diagrams.]{%
The graph operation $\ko{i}$ maps an $\nkmhv{\hel}$ twistor diagram
into an $\nkmhv{\hel+1}$ twistor diagram as shown in the top row.
On Landau diagrams, this corresponds to replacing
 a massless corner by a massive corner;
such an operation is effectively an inverse collinear limit.
The shaded region in the figures
represents an arbitrary planar sub-graph.
A dashed external line on a Landau diagram may be either one
massless external leg so the whole corner is massive,
or completely removed so the whole corner is massless.
}
 \label{fig:graph-operation}
\end{figure}
%
%

The first graph operation $\ko{}$
increments the helicity of the diagram on which it operates.
(The name $\ko{}$ is a reminder that it increases $\hel$.)
Its operation is demonstrated in \figRef{graph-operation}.
Specifically, $\ko{i}$ replaces a filled node
at a point $i$ along $\mathcal{L}$ by two empty
nodes, one at $i$ and a second one on a new non-MHV intersection
added to the diagram.
Since $\nFull$ decreases by one but $\nEmpty$ increases by two under this operation,
it is clear from~\eqnRef{k-from-mts} that $\ko{i}$ always
increases by one the minimal value of $\hel$ on which the branch
indicated by the twistor diagram has support.
From the point of view of Landau diagrams,
this operation replaces a massless node with a massive one,
as illustrated in the bottom row of~\figRef{graph-operation},
and hence it may be viewed as an ``inverse'' collinear limit.

The other two graph operations $\ro{}$ and $\uo{}$ both correspond
to relaxations, as defined in~\secRef{boundaries}, since
they each reduce the number of on-shell conditions by one, stepping
thereby one column to the right in~\figRef{one-loop-graph-flow}.

\begin{figure}
\centering
\includegraphics{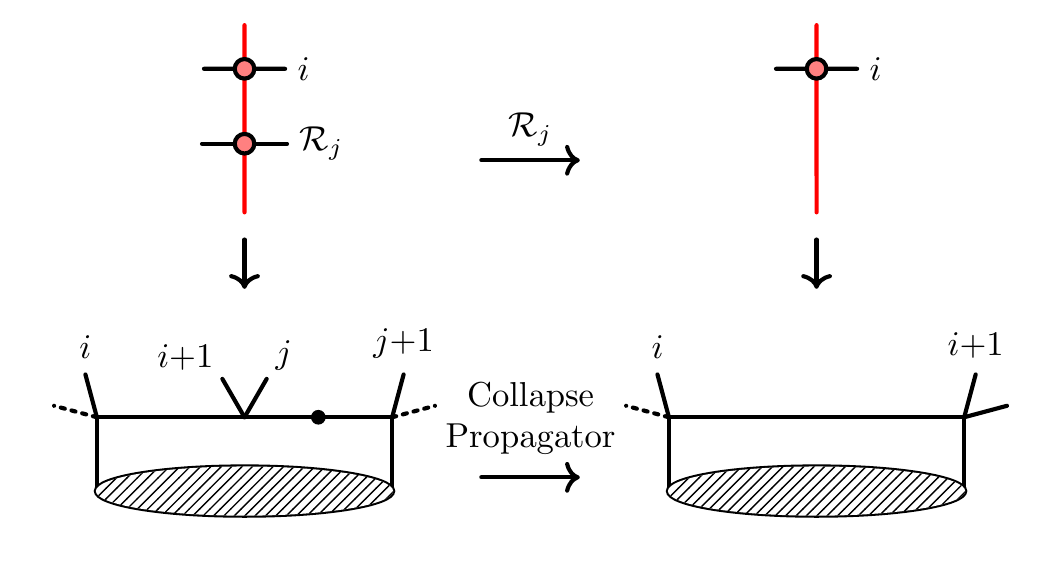}
\caption[Removing graph operation on momentum twistor diagrams.]{%
The graph operation $\ro{j}$ relaxes $\mathcal{L}$
by removing the condition
that $\mathcal{L}$ must pass through the line $(j\,j{+}1)$;
this is equivalent to removing the on-shell condition
$\langle \mathcal{L} \, j \, j{+}1 \rangle =0$.
On Landau diagrams, this corresponds to collapsing the propagator
indicated by the filled dot in the bottom figure on the left.
The shaded region in the figures
represents an arbitrary planar sub-graph.
A dashed external line on a Landau diagram may be either one
massless external leg so the whole corner is massive,
or completely removed so the whole corner is massless.
It is to be understood that the graphical notation implies
that $j \ne i+2$ and
$i \ne j+2$; otherwise, the two empty nodes in the top left
diagram would be represented by a single filled node on
which the action of $\ro{}$ is undefined; the appropriate
graph operation in this case would instead be $\uo{}$.
}
 \label{fig:remove-go}
\end{figure}

The operation $\ro{i}$ simply removes
(hence the name $\ro{}$) an empty node $i$ from a twistor diagram,
as shown in~\figRef{remove-go}.
This corresponds to removing
$\langle \mathcal{L} \, i \, i+1 \rangle=0$
from the set of on-shell conditions satisfied by
$\mathcal{L}$\footnote{Note that in line with the conventions
adopted in~\secRef{graphical-notation} we
label $\ro{i}$ only with the
smaller label of a pair $(i\,i{+}1)$.}.

\begin{figure}
\centering
\includegraphics{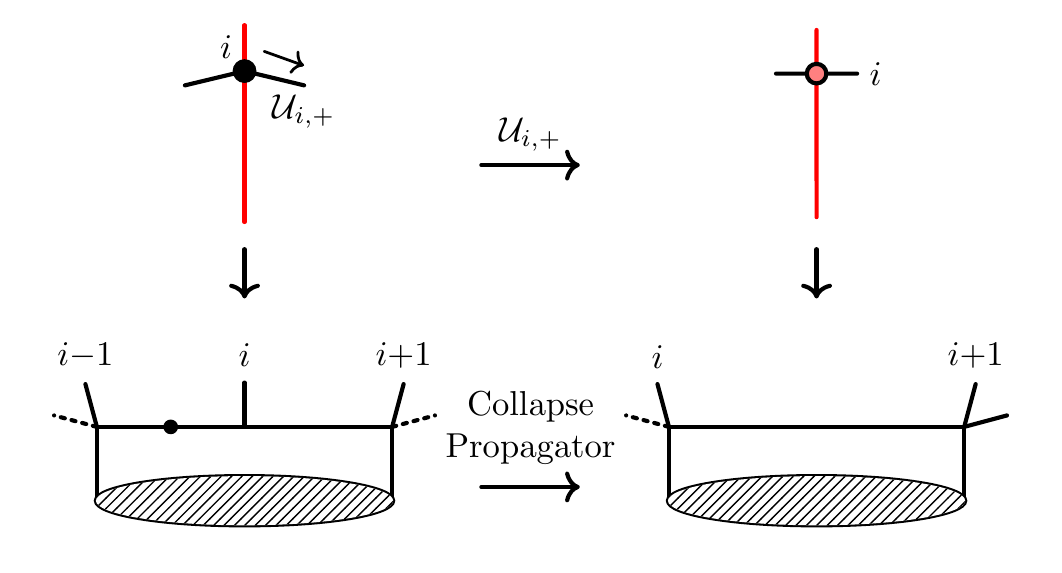}
\caption[Un-pinning graph operation on momentum twistor diagrams.]{%
The graph operation $\uo{i,+}$ relaxes a line $\mathcal{L}$
constrained to
pass through the point $i$,
shifting it to lie only along
the line $(i\,i{+}1)$.
This is equivalent to removing the on-shell
constraint $\langle \mathcal{L} \, i{-}1 \, i  \rangle = 0$.
(The equally valid relaxation $\uo{i,-}$, not pictured
here, lets the intersection point slide onto $(i{-}1\,i)$.)
On Landau diagrams, this corresponds to collapsing the propagator
indicated by the filled dot in the bottom figure on the left.
The shaded region in the figures
represents an arbitrary planar sub-graph.
A dashed external line on a Landau diagram may be either one
massless external leg so the whole corner is massive,
or completely removed so the whole corner is massless.
As explained in the caption of~\figRef{remove-go}, the
$\uo{}$ operation can be thought of as a special
case of the $\ro{}$ operation, and we distinguish the two because
only the
latter can change the helicity sector $\hel$.
}
 \label{fig:unpin-go}
\end{figure}

The last operation, $\uo{}$, corresponds to ``un-pinning''
a filled node (hence ``$\uo{}$'').
Un-pinning means removing
one constraint from a pair $\langle \mathcal{L}\,i{-}1\,i\rangle
= \langle \mathcal{L}\,i\,i{+}1\rangle = 0$.
The line $\mathcal{L}$, which was pinned to the point $i$, is then
free to slide along the line segment $(i{-}1\,i)$ or $(i\,i{+}1)$
(for $\uo{i,-}$ or $\uo{i,+}$, respectively).
In the twistor diagram, this is depicted by replacing the filled
node at the point $i$ with a single
empty node along the line segment $(i\,i{\pm}1)$ (see~\figRef{unpin-go}).
Only $\uo{+}$ appears in~\figRef{one-loop-graph-flow} because
at one loop, all diagrams generated by any $\uo{-}$ operation are equivalent,
up to relabeling, to some diagram generated by a $\uo{+}$.
In general, however, it is necessary to track the subscript $\pm$
since both choices are equally valid relaxations and can
yield inequivalent twistor and Landau diagrams.
From \figRef{one-loop-graph-flow}, we read off
the following identity among the operators
acting on any diagram $g$:
\begin{align}
\uo{j,+} g = \ro{k} \ko{j,+} g\,.
\end{align}

There was no reason to expect the
simple graphical pattern of~\figRef{unpin-go}
to emerge among the twistor diagrams.
Indeed in~\secRef{one-loop-branches}
we simply listed all possible sets of on-shell conditions without taking such an organizational
principle into account.  At higher loop order, however, the problem
of enumerating all boundaries of
$\nkmhv{\hel}$ amplituhedra benefits
greatly from the fact that all valid configurations of each single
loop can be iteratively generated via these simple rules,
starting from the maximal
codimension MHV boundaries.
Stated somewhat more abstractly, these graph operations are instructions
for naturally associating boundaries of different amplituhedra.

%
%
%
%

%
%
%
%

Before concluding this section it is worth noting
(as is evident in~\figRef{one-loop-graph-flow})
that relaxing a low-$\hel$ boundary
can never raise the minimum value of
$\hel$ for which that type of boundary is valid.
In other words, we find that if
$\mathcal{A}_{n,\hel,1}$ has a boundary of type $B$,
and if $B'$ is a relaxation of $B$, then $\mathcal{A}_{n,\hel,1}$
also has boundaries of type $B'$.
This property does not hold in general beyond one loop;
a counterexample involving two-loop MHV amplitudes
appears in Fig.~4 of~\cite{Dennen:2016mdk}.

\section{Solving Landau Equations in Momentum Twistor Space}
\label{sec:momentumtwistorlandauequations}

As emphasized in~\secRef{summary}, the Landau equations
naturally associate to each boundary of an amplituhedron
a locus in $\Conf_n(\mathbb{P}^3)$ on which the corresponding
amplitude has a singularity.
In this section we review the results of solving the Landau
equations for each of the one-loop branches classified
in~\secRef{one-loop-branches}, thereby carrying
out step~2 of the algorithm summarized in~\secRef{summary}.
The results of this section were already tabulated in~\cite{Dennen:2015bet},
but we revisit the analysis, choosing just two examples, in order
to demonstrate the simplicity and efficiency
of these calculations when carried out directly in momentum
twistor space.
The utility of this method is on better display
in the higher-loop examples to be considered in the sequel.

As a first example, we consider the tadpole on-shell condition
\begin{align}
\label{eqn:one}
f_1 \equiv \langle \mathcal{L}\,i\,i{+}1\rangle = 0\,.
\end{align}
We choose any two other points
$Z_j$, $Z_k$ (which generically satisfy $\langle i\,i{+}1\,j\,k\rangle \ne 0$)
in terms of which to parameterize
\begin{align}
\label{eqn:eight}
\mathcal{L} = (Z_i + d_1 Z_j + d_2 Z_k, Z_{i+1} + d_3 Z_j + d_4 Z_k)\,.
\end{align}
Then the on-shell
condition~(\ref{eqn:one}) admits solutions when
\begin{align}
\label{eqn:five}
d_1 d_4 - d_2 d_3 = 0\,,
\end{align}
while the four Kirchhoff conditions~(\ref{eqn:landaukirchhoffequations}) are
\begin{align}
\label{eqn:four}
\alpha_1 d_4 = - \alpha_1 d_3 = - \alpha_1 d_2 = \alpha_1 d_1 = 0\,.
\end{align}
The
only nontrivial solution
(that means $\alpha_1 \ne 0$; see~\secRef{landauequations})
to the equations~(\ref{eqn:five}) and~(\ref{eqn:four})
is to set all four $d_A = 0$.
Since this solution exists for all (generic)
projected external data, it does not correspond to a branch point
of an amplitude and is uninteresting to us.
In other words, in this case the locus we associate to a boundary
of this type is all of $\Conf_n(\mathbb{P}^3)$.

As a second example, consider the
two on-shell conditions corresponding to the two-mass
bubble
\begin{align}
\label{eqn:three}
f_1 \equiv \langle \mathcal{L}\,i\,i{+}1\rangle = 0\,, \qquad
f_2 \equiv \langle \mathcal{L}\,j\,j{+}1\rangle = 0\,.
\end{align}
In this case a convenient parameterization is
\begin{align}
\label{eqn:ten}
\mathcal{L} = (Z_i + d_1 Z_{i+1} + d_2 Z_k,
Z_j + d_3 Z_{i+1} + d_4 Z_k)\,.
\end{align}
Note that an asymmetry between $i$ and $j$ is necessarily
introduced because we should not allow more than four distinct
momentum twistors to appear in the parameterization, since
they would necessarily be linearly dependent, and
we assume of course that $Z_k$ is generic
(meaning, as before, that $\langle i\,i{+}1\,j\,k\rangle \ne 0$).
Then
\begin{align}
f_1 &= - d_2 \langle i\, i{+}1\, j\, k\rangle\,,\cr
f_2 &= d_3 \langle i\, i{+}1\, j\, j{+}1 \rangle
+ d_4 \langle i\, j\, j{+}1\, k\rangle
+ (d_1 d_4 - d_2 d_3) \langle i{+}1\,j\,j{+}1\,k\rangle
\end{align}
and the Kirchhoff conditions are
\begin{align}
\label{eqn:nine}
\left(\begin{matrix}
0 & d_4 \langle i{+}1\,j\,j{+}1\,k\rangle \\
- \langle i\,i{+}1\,j\,k\rangle & - d_3 \langle i{+}1\,j\,j{+}1\,k\rangle \\
0 & \langle i\,i{+}1\,j\,j{+}1 \rangle-d_2 \langle i{+}1\,j\,j{+}1\,k\rangle\\
0 & \langle i\,j\,j{+}1\,k\rangle + d_1 \langle i{+}1\,j\,j{+}1\,k\rangle
\end{matrix}
\right)
\left(
\begin{matrix}
\alpha_1\\
\alpha_2 \end{matrix}\right)=0\,.
\end{align}
Nontrivial solutions exist only if all $2 \times 2$ minors
of the $4 \times 2$ coefficient matrix vanish.
Three minors are trivially zero, and the one computed
from the second and third rows evaluates
simply to
\begin{align}
- \langle i\,i{+}1\,j\,k\rangle \langle i\,i{+}1\,j\,j{+}1\rangle = 0
\end{align}
using the on-shell condition $f_1 = - d_2 \langle i\,i{+}1\,j\,k\rangle = 0$.
If this quantity vanishes, then the four remaining
constraints (the two on-shell conditions $f_1 = f_2 = 0$ and
the two remaining minors) can be solved for the four $d_A$,
and then~\eqnRef{nine} can be solved to find the two $\alpha_J$'s.
Since $\langle i\,i{+}1\,j\,k \rangle \ne 0$ by assumption,
we conclude that the Landau equations admit nontrivial solutions
only on the codimension-one locus in $\Conf_n(\mathbb{P}^3)$
where
\begin{align}
\label{eqn:six}
\langle i\,i{+}1\,j\,j{+}1\rangle = 0\,.
\end{align}

These two examples demonstrate that in some cases
(e.g. the tadpole example) the Landau equations admit solutions
for any (projected) external data,
while in other cases (e.g. the bubble example)
the Landau equations admit solutions
only when there is a codimension-one constraint
on the external data.
A common feature of these examples
is that some care must be taken in choosing how to parameterize
$\mathcal{L}$.
In particular, one must never express
$\mathcal{L}$ in terms of four momentum twistors
($Z_i$, $Z_j$, etc.) that appear in the specification
of the on-shell conditions; otherwise, it can be impossible
to disentangle the competing
requirements that these satisfy
some genericity (such as $\langle i\,i{+}1\,j\,k\rangle \ne 0$
in the above examples)
while simultaneously hoping to tease out the constraints they
must satisfy in order to have a solution
(such as~\eqnRef{six}).
For example, although one might have been tempted
to preserve the symmetry between $i$ and $j$,
it would have been a mistake to use
the four twistors $Z_i$, $Z_{i+1}$, $Z_j$ and $Z_{j+1}$
in~\eqnRef{ten}.

Instead, it is safest to always pick four completely generic points
$Z_a, \ldots, Z_d$ in terms of which to parameterize
\begin{align}
\mathcal{L} =\left( \begin{matrix}
1 & 0 & d_1 & d_2 \cr
0 & 1 & d_3 & d_4 \end{matrix} \right)
\left( \begin{matrix}
Z_a \cr
Z_b \cr
Z_c \cr
Z_d \end{matrix}\right).
\label{eqn:seven}
\end{align}
The disadvantage of being so careful is that intermediate steps
in the calculation become much more lengthy, a problem we avoid
in practice by using a computer algebra system such as Mathematica.

The results of this analysis
for all one-loop branches are summarized in~\tabRef{bigtable}.
Naturally these are in accord with those of~\cite{Landau:1959fi}
(as tabulated in~\cite{Dennen:2015bet}).
At one loop it happens that the singularity locus is the same for each branch
of solutions to a given set of on-shell conditions, but this
is not generally true at higher loop order.

\section{Singularities and Symbology}
\label{sec:symbology}

As suggested in the introduction (and explicit even in the title
of this paper), one of the goals of our research program
is to provide a priori derivations of
the~\emph{symbol alphabets}
of various amplitudes.
We refer the reader
to~\cite{Goncharov:2010jf} for more details, pausing
only to recall that the symbol alphabet of a generalized
polylogarithm function
$F$ is a finite list of \emph{symbol letters}
$\{ z_1, \ldots, z_r\}$
such that $F$ has \emph{logarithmic}
branch cuts (i.e., the cover has infinitely many
sheets)\footnote{These branch cuts
usually do not all live on the same sheet; the symbol alphabet
provides a list of all branch cuts that can be accessed
after analytically continuing $F$ to arbitrary sheets.} between
$z_i = 0$ and $z_i = \infty$ for each $i=1,\ldots,r$.

To date, symbol alphabets have been determined by explicit computation
only for two-loop MHV amplitudes~\cite{CaronHuot:2011ky};
all other results on multi-loop SYM amplitudes
in the literature are based on a conjectured
extrapolation of these results to higher loop order.
Throughout the paper we have however been careful to phrase
our results in terms of branch points, rather than symbol
letters, for two reasons.

First of all, amplitudes in SYM
theory are expected to be expressible as generalized polylogarithm
functions, with symbol letters that have a familiar structure
like those of the entries in the last column of \tabRef{bigtable},
only for sufficiently low (or, by parity conjugation, high) helicity.
In contrast, the Landau equations
are capable of detecting branch points
of even more complicated amplitudes,
such as those containing elliptic polylogarithms,
which do not have traditional symbols\footnote{It
would be interesting to understand
how the ``generalized symbols" of such amplitudes
capture the singularity loci revealed by the
Landau equations.}.

Second, even for amplitudes which do have symbols, determining the
actual symbol alphabet from the singularity loci of the amplitude
may require nontrivial extrapolation.
Suppose that the Landau equations reveal that some amplitude
has a branch point at $z=0$ (where, for example, $z$ may be one of the
quantities in the last column of~\tabRef{bigtable}).
Then the symbol alphabet should contain a letter $f(z)$, where $f$ in
general could be an arbitrary function of $z$, with
branch points arising in two possible ways.
If $f(0) = 0$, then the amplitude will have a logarithmic branch
point at $z=0$~\cite{Maldacena:2015iua}, but
even if $f(0) \ne 0$, the amplitude can have an
\emph{algebraic} branch point
(so the cover has finitely many sheets) at $z=0$ if
$f(z)$ has such a branch point there.

We can explore this second notion empirically
since all one-loop amplitudes in SYM theory, and in particular
their symbol alphabets, are well-known
(following from
one-loop integrated amplitudes in for example,%
~\cite{tHooft:1978jhc,Bern:1993kr,Bern:1994zx,
Bern:1994ju,Brandhuber:2004yw,Bern:2004ky,Britto:2004nc,Bern:2004bt,Ellis:2007qk}).
According to our results from~\tabRef{bigtable}, we find that
one-loop amplitudes only have branch points on loci of the form
\begin{itemize}
\item
 $\langle i\,i{+}1\,j\,j{+}1\rangle = 0$
or $\langle i\,\overline{j} \rangle = 0$ for $0 \le \hel \le n-4$,
\item
$\langle i(i{-}1\,i{+}1)(j\,j{+}1)(k\,k{+}1)\rangle = 0$ for $1 \le \hel \le n-5$, and
\item
$\Delta_{ijk\ell} = 0$ (defined in~\tabRef{bigtable}) for
$2 \le \hel \le n-6$,
\end{itemize}
where $i,j,k,\ell$ can all range from $1$ to $n$.
Happily, the first two of these are in complete accord with the symbol letters
of one-loop MHV and NMHV amplitudes,
but the third reveals the foreshadowed
algebraic branching
since $\Delta_{ijk\ell}$ is not a symbol letter
of the four-mass box integral contribution to $\nkmhv{2 \le \hel \le n-6}$
amplitudes. Rather, the symbol alphabet of this amplitude consists
of quantities of the form
\begin{align}
f_{ij} \equiv \langle i\,i{+}1\,j\,j{+}1\rangle
\quad \textrm{and} \quad
f_{i\ell}f_{jk} \pm( f_{ik} f_{j\ell}- f_{ij}
f_{k\ell}) \pm \sqrt{\Delta_{ijk\ell}}\,,
\label{eqn:algebraic}
\end{align}
where the signs may be chosen independently.
Since no symbol letter vanishes on the locus $\Delta_{ijk\ell} = 0$,
amplitudes evidently do not have logarithmic branch points
on this locus. Yet it is evident from the second
expression of~(\ref{eqn:algebraic})
that amplitudes with these letters have algebraic
(in this instance, square-root- or double-sheet-type)
branch points when $\Delta_{ijk\ell} = 0$.

Although we have only commented on the structure
of various potential symbol entries and branch point loci here,
let us emphasize that the methods of this paper can be used to
determine precisely which symbol entries can appear
in any given amplitude.
For example, \tabRef{bigtable} can be used
to determine values of $i$, $j$ and $k$ for which the
letter $\langle i(i{-}1\,i{+}1)(j\,j{+}1)(k\,k{+}1)\rangle$
can appear, as well as in which one-loop amplitudes,
indexed by $n$ and $k$, such letters will appear.
An example of a fine detail along these lines evident
already in~\tabRef{bigtable} is the fact that
all NMHV amplitudes have branch points of two-mass easy type
except for the special case $n=6$, in accord
with Eq.~(2.7) of~\cite{Kosower:2010yk}.

We conclude this section by remarking that the problem of
deriving symbol alphabets from the Landau singularity loci may remain
complicated in general, but we hope
that the simple, direct correspondence
we have observed for certain one-loop amplitudes (and which was also observed
for the two-loop MHV amplitudes studied
in~\cite{Dennen:2016mdk}) will continue to hold
at arbitrary loop order for sufficiently simple singularities.

\section{Conclusion}

This paper presents first steps down the path of understanding
the branch cut structure of SYM amplitudes for general helicity,
following the lead of~\cite{Dennen:2016mdk} and using
the recent ``unwound'' formulation of the
amplituhedron from~\cite{Arkani-Hamed:2017vfh}.
Our algorithm is conceptually simple: we first enumerate the boundaries
of an amplituhedron, and from there, without resorting
to integral representations, we use the Landau equations directly
to determine the locations of branch points of the corresponding amplitude.

One might worry that each of these steps grows rapidly
in computational complexity at higher loop order.
Classifying boundaries of amplituhedra is
on its own
a highly nontrivial
problem, aspects of which have been
explored in~\cite{Arkani-Hamed:2013kca,Franco:2014csa,Bai:2015qoa,Galloni:2016iuj,Karp:2017ouj}.
In that light, the graphical tools presented
in~\secRef{graphical-organization},
while already useful for organizing results as in~\figRef{one-loop-graph-flow},
hint at the more enticing possibility of
a method to enumerate
twistor diagrams corresponding to
all $\mathcal{L}$-boundaries
of any given $\mathcal{A}_{n,\hel,L}$.
Such an algorithm would start with the maximal codimension twistor diagrams
at a given loop order, and apply
the operators of~\secRef{graphical-organization} in all ways until no further operations
are possible.  From these twistor diagrams come Landau diagrams,
and from these come the branch points via the Landau equations.
We saw in~\cite{Dennen:2016mdk}
and~\secRef{momentumtwistorlandauequations} that analyzing
the Landau equations can be made very simple in momentum twistor space.

Configurations of loop momenta in (the closure of) MHV amplituhedra
are represented by non-negative $D$-matrices.
In general, non-MHV configurations must be represented by
indefinite $D$-matrices, but we observed
in~\secRef{emergent} that even
for non-MHV
amplituhedra,
$D$ may always be chosen
non-negative for all configurations on
$\mathcal{L}$-boundaries.
This `emergent positivity' plays a crucial role by allowing
the one-loop $D$-matrices presented in
Secs.~\ref{sec:mhvlower},
\ref{sec:nmhvlower} and~\ref{sec:nnmhvlower} to be trivially recycled
at higher values of helicity.
One way to think about this is to say that going beyond MHV level
introduces the $C$-matrix which ``opens up''
additional configuration space in which an otherwise indefinite
$D$-matrix can become positive.

While the one-loop all-helicity results we obtain are interesting in their own
right as first instances of all-helicity statements,
this collection of information is valuable because it
provides the building blocks for the two-loop analysis in the sequel.
There we will argue that the two-loop twistor diagrams with helicity $\hel$
can be viewed as compositions
of two one-loop diagrams with
helicities $\hel_1$ and $\hel_2$ satisfying $\hel = \hel_1+\hel_2$ or
$\hel_1+\hel_2+1$.
We will also explore in detail the relation to on-shell diagrams,
which are simply Landau diagrams with
decorated nodes.

More speculatively, the ideas
that higher-loop amplitudes can be constructed from lower-loop amplitudes, and
that there is a close relation to on-shell diagrams,
suggests the possibility that this
toolbox may also be useful for finding symbols in the full, nonplanar
SYM theory.
For example, enumerating the on-shell conditions as we do here in the planar
sector is similar in spirit to the nonplanar examples of~\cite{Bern:2015ple}
where certain integral representations
were found such that individual integrals had support on only
certain branches\footnote{Already in the planar case, one might
interpret our algorithm as applying the Landau equations
to integrands constructed in expansions
around boundaries of amplituhedra, which is reminiscent of the
prescriptive unitarity
of~\cite{Bourjaily:2017wjl}.}.
There are of course far fewer known results in the nonplanar SYM theory,
though there have been some preliminary studies%
~\cite{Arkani-Hamed:2014via,Bern:2014kca,Bern:2017gdk,Franco:2015rma,Bourjaily:2016mnp}.

\acknowledgments

We have benefited greatly from
very stimulating discussions with N.~Arkani-Hamed.
This work was supported in part by: the US Department of Energy under contract
DE-SC0010010 Task A,
Simons Investigator Award \#376208 (JS, AV),
the Simons Fellowship Program in Theoretical Physics (MS),
the IBM Einstein Fellowship (AV),
the National Science Foundation under Grant No. NSF PHY-1125915 (JS),
and the Munich Institute for Astro- and Particle
Physics (MIAPP) of the DFG cluster of excellence ``Origin and Structure of the
Universe'' (JS).
MS and AV are also grateful to the CERN theory group for hospitality
and support during the course of this work.

\end{document}